# Mapping the Landscape of Affective Extended Reality: A Scoping Review of Biodata-Driven Systems for Understanding and Sharing Emotions

Zhidian Lin
School of Computing Technologies
RMIT University
Melbourne, Australia
zhidian.lin@student.rmit.edu.au

Allison Jing
School of Computing Technologies
RMIT University
Melbourne, Australia
allison.jing@rmit.edu.au

Ziyuan Qu
School of Design
RMIT University
Melbourne, Australia
ziyuan.qu@student.rmit.edu.au

Fabio Zambetta
School of Computing Technologies
RMIT University
Melbourne, Australia
fabio.zambetta@rmit.edu.au

Ryan M. Kelly
School of Computing Technologies
RMIT University
Melbourne, Australia
ryan.kelly@rmit.edu.au

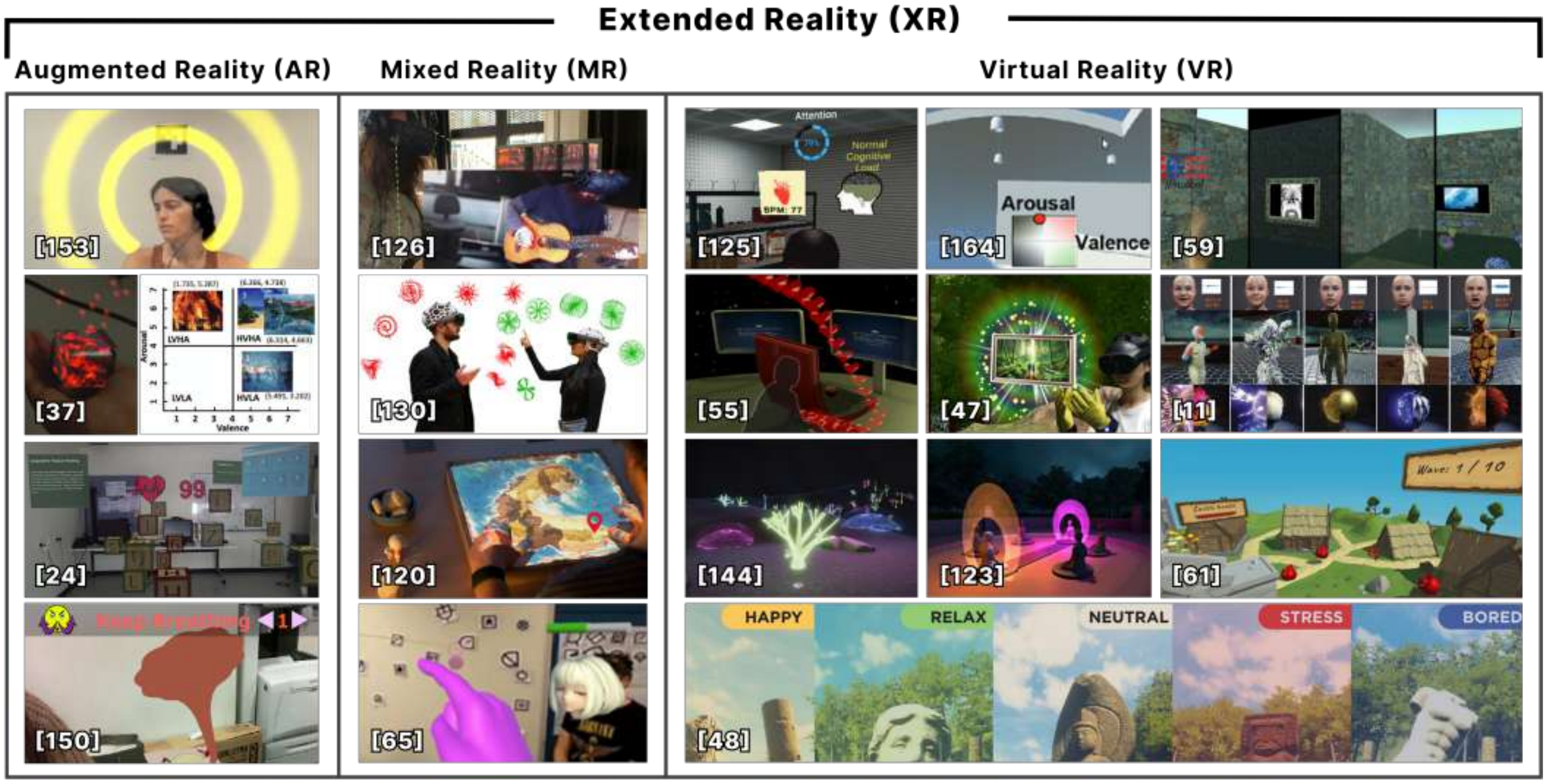

**Figure 1: Examples of affective XR systems, with reference numbers indicated in the bottom-left corner. From left to right, AR/MR: Limited virtual immersion to amplify emotions by augmenting visual stimuli on either human subjects, physical objects, or environmental spaces. VR: Full virtual immersion to manifest emotions by providing different affordances with multimodal feedback (e.g. visual, auditory, and haptic) in fully controlled virtual environments.**

## Abstract

This paper introduces the notion of affective extended reality (XR) to characterise XR systems that use biodata to enable understanding of emotions. The HCI literature contains many such systems, but they have not yet been mapped into a coherent whole. To address this, we conducted a scoping review of 82 papers that explore the nexus of biodata, emotions, and XR. We analyse the technologies used in these systems, the interaction techniques employed, and the methods used to evaluate their effectiveness. Through our analysis, we contribute a mapping of the current landscape of affective XR, revealing diversity in the goals for enabling emotion sharing. We demonstrate how HCI researchers have explored the design of the interaction flows in XR biofeedback systems, highlighting key design dimensions and challenges in understanding emotions. We discuss underused approaches for emotion sharing and highlight opportunities for future research on affective XR.

## 1 Introduction

Emotions are a fundamental part of human experience. Every day, people continually express their emotions and interpret those of others in ways that significantly shape social interactions [41, 122]. Although face-to-face communication facilitates intuitive emotional expression, the rapid proliferation of digital technology has transformed our ability to enhance the way emotions are shared between people [64]. These applications span medical and healthcare settings [35, 58, 127], gaming and entertainment [1, 55, 97, 142], training environments [9, 39, 98], education [99, 146, 164], and collaborative work [52, 67]. A substantial number of systems designed to support emotion sharing in these contexts employ extended reality (XR) technologies, including virtual reality (VR), augmented reality (AR), and mixed reality (MR), to create powerful experiences centred around interpersonal awareness and empathy [64].

These emotion-sharing XR systems often involve the collection and use of biodata to offer more objective [49, 130] and reliable [28] methods for emotion communication. Researchers have presented systems that use biodata to evoke emotional responses in virtual environments [5, 31, 38, 76], measure affective states such as anxiety [74, 96, 132], and infer user preferences for content customisation based on emotional profiles [8, 19, 63, 101]. As shown in Figure 1, the HCI community has played a leading role in this work, with notable contributions including emotion-augmented games [59, 61], empathy-supporting communication systems [130, 153], and technologies that enable relaxation through biofeedback and meditative breathwork [124, 150]. However, this growing body of literature remains fragmented across multiple domains and publication venues. This fragmentation leaves designers without systematic guidance on emotion representation strategies, potentially contributing to ineffective implementations. Researchers also risk reinventing approaches without understanding which work in specific contexts. No work has yet synthesised the literature to identify common design considerations, challenges, and directions for future research.

In this paper, we introduce the notion of 'affective XR' to characterise and systematically examine these systems as a coherent body of work. We define affective XR systems as those that collect biodata from one or more users, transform this data into meaningful representations of emotion through biofeedback, and display this feedback to users to influence their behaviour and experiences. This process represents a fundamentally creative design endeavour requiring a variety of choices: navigating diverse biodata modalities from heart rate variability [9, 87, 152] to facial expressions [51], selecting from an ever-expanding palette of XR technologies, and considering varied approaches for conveying emotional understanding through options such as abstract visualisations [35, 130], embodied avatars [123, 137], or environmental modifications [48, 49]. These multifaceted design decisions raise significant questions about how to facilitate emotion sharing between users, and how different design approaches might yield varying outcomes for emotional communication and behavioural influence.

We report findings from a scoping review of 82 papers examining affective XR technologies across HCI venues. Our review makes three contributions to the HCI literature. First, we conduct an investigation of the landscape of affective XR, examining how biodata serves as input for emotion understanding and sharing in XR systems. Second, we develop a taxonomy outlining biodata flows in affective XR technologies, and identify a set of nine design dimensions for representing emotions along with four key design challenges. Lastly, we discuss future research directions for biodata-driven affective XR, contributing to emotion research in HCI [156] and the growing conversation around the use of physiological data within the CHI community [20, 94].

## 2 Background

This section will first clarify the definition of the key terminologies used in the following sections, then review existing research and gaps to motivate our work.

### 2.1 Relevant Terminology and Scope

Affective computing, a field formally established by Picard's seminal work in 2000, defines three core functions: recognising, representing, and simulating human emotion [105]. While significant progress has since been made in empathic design and emotionally aware systems [88, 129], the application of affective computing principles within immersive virtual environments is still a nascent area of research. To explore this emerging intersection, this review introduces the concept of affective XR, which we define as the use of immersive technologies, including virtual reality (VR), augmented reality (AR), and mixed reality (MR), to support affective understanding, social interaction and communication, and enhance emotional experience through a contextual embodied head-mounted display.

Emotion modelling in affective computing commonly draws upon two major theoretical frameworks. The first, Ekman's *Discrete Model*, posits six basic and universal emotions: fear, anger, joy, sadness, disgust, and surprise [36]. The second, Russell's *Dimensional*

*Model*, conceptualises emotions within a continuous space defined by arousal (activation, i.e., low/high) and valence (pleasantness, i.e., negative/positive) [121]. A central distinction between these models lies in their representational approaches: the Discrete Model categorises emotions into distinct classes, whereas the Dimensional Model treats emotions as gradable experiences along scalable continua, which are not adequately captured by a finite set of discrete categories [6, 50]. This review, however, intentionally avoids adopting or evaluating specific emotion models. Instead, we focus on reporting the particular emotional states, such as joy, fear, calm, or arousal, valence, as presented in the papers describing affective XR systems. This emphasis on tangible emotional vocabulary rather than model-specific terminology allows us to foreground concrete affective interactions and applications, prioritising real-world emotional experiences over theoretical classifications.

Biosignals encompass a wide range of continuous physiological phenomena that reflect an organism's internal state [103]. In this paper, we adopt the broader term 'biodata' to refer to biofeedback within XR systems. Biodata, including heart rate, breath rate, muscle tension, electrodermal activity, and brain activity, provides an objective [49, 130] and unobtrusive [80] means of inferring affective states, circumventing biases inherent in self-report methods [78, 94, 134]. An advantage of physiological responses is that they can be dynamically integrated as an implicit input to adaptive XR applications, enabling real-time system adaptation based on the user's state [115]. For instance, elevated heart rate may signal anxiety in therapeutic VR settings [69] or excitement in immersive gaming contexts [70]. Biodata can thus prompt system modifications to foster desired states, such as supporting emotion regulation from stress to relaxation [152, 157], or facilitate responsive adaptations to individual physiological conditions [19, 143, 147].

## 2.2 Existing Research and Gaps

Our review aims to organise the XR landscape by adopting a distinct focus on biodata-driven emotional interfaces. Importantly, we take a meaningfully different perspective from other recent reviews of the XR literature. One review by Jinan et al. [64] systematically reviewed immersive technologies for empathic computing, including VR, AR and MR. However, their work adopted a broader perspective that did not specifically foreground biodata as an input modality, and included studies that fall outside the boundaries of affective XR. These include studies on virtual pedagogical agents designed to elicit emotional responses [13], critical examinations of VR as an 'Empathy Machine' for emotional storytelling [12], AR simulations of sensory impairment to evoke emotional awareness [44], and animal embodiment experiences aimed at generating empathic connections [108]—none of which rely on biodata.

Separately, other reviews offer broader examinations of biodata systems without a specific emphasis on XR technologies [42, 95, 165]. For example, Moge et al. [95] conducted a comprehensive survey on the social sharing of physiological data, developing a conceptual framework centred on symmetry in biodata access and representation. Their review spans a wide range of non-XR systems focused on emotional communication, including public installations that broadcast heart sounds to create intimate emotional experiences [57], wearables that communicate breathing patterns to foster emotional connection [40], mobile messaging apps that integrate heart rate sharing to support emotional awareness [54], and smartwatch applications that enable mood-based interaction for lightweight emotional expression [84]. While their symmetry framework offers valuable insights into social biofeedback design for emotional communication, their review is constrained by an exclusive focus on multi-user empathy scenarios, omitting single-user contexts where biosignals are interpreted through emotional self-awareness and contextual cues [138]. Furthermore, their work does not specifically address how XR environments uniquely afford embodied emotional experiences and empathic engagement, thereby leaving a gap in understanding how immersive technologies can enhance emotional communication through integrated biosignal interfaces.

In contrast, our review examines the role of biodata as an input modality for emotion recognition and empathic interaction within XR environments. Complementing this input, XR technologies provide a distinct set of opportunities for emotion representation compared to conventional interfaces such as desktop dashboards [162], mobile apps [54] or smartwatches [84]. AR/MR devices have the potential to amplify feelings by augmenting visual stimuli in the real world, including people, objects and the environment. Conversely, VR can enable manipulation of emotion within a fully controlled virtual environment, providing opportunities to manifest emotions through interacting with objects, exploring virtual worlds, or even reconfiguring physics to induce affect. By focusing on these unique affordances of XR, we extend Moge et al.'s symmetry framework [95] to encompass both single-user emotional self-regulation and multi-user systems. We incorporate discussions such as emotional visualisations [106], auditory and haptic cues for empathic communication [100], which are critical for supporting emotional understanding and empathy in immersive environments. Thus, our work narrows the technological scope to XR systems while broadening the conceptual lens to include both individual emotional awareness and social empathic interaction use cases among multiple users, offering a dedicated analysis of how biodata is sensed, processed, and rendered to support emotional understanding and empathic engagement in XR environments.

# 3 Method

We conducted a scoping review to understand the current landscape of affective XR. We focused our review on XR systems that first collect biodata to infer emotions before conveying a representation of these emotions back to one or more users. Our review followed the established scoping review process outlined by Arksey and O'Malley [2], which has been used widely by the HCI community [e.g. 140, 160]. We chose the scoping review approach because we wanted to understand the current scope of affective XR research while identifying challenges and opportunities for future work.

## 3.1 Developing the Research Questions

Our review has the broad aim of mapping the landscape of affective XR systems. However, 'affective XR' is a new term that we are using to characterise the existing literature. Therefore, we began with a broad research question that guided our review process: *How have XR systems been designed to support emotion sharing?*

This initial question was refined as our focus shifted toward the use of biodata, because we discovered that biodata was used in a significant number of papers describing affective XR systems. This led us to focus on more specific questions, including: *How has biodata been used to enable emotion sharing in XR? What are the motivations for representing emotions via biodata?*

From exploring these initial questions, we arrived at two main research questions that we answer through our scoping review:

**RQ1:** How have researchers designed XR systems to enable emotion understanding and sharing?

**RQ2:** What are the key challenges when enabling emotion understanding and sharing in XR environments?

## 3.2 Identifying Articles

We conducted our searches from May to June 2025. Our search process began with initial pilot searches of the ACM Digital Library and IEEE Xplore to examine the quantity of research published on emotions and XR. After obtaining promising results, we developed a search string to identify articles at the nexus of biodata, emotion, and XR technologies. We developed our search string by beginning with keywords used in other recent literature reviews. These included keywords from reviews of biodata and physiological signals [14, 46, 102, 158], a short set of emotion and affective-centric keywords from reviews of empathic computing [64, 122, 139], and a comprehensive set of XR-related keywords from a recent HCI review of AI and XR by Hirzle et al. [56]. We added additional keywords to make our search as comprehensive as possible. These included general terms like "mood" or synonyms such as "bio-signal", along with keywords to capture specific biodata, including "ECG", "skin temperature" and "brain activity". Appendix A provides our complete search string.

We used the search string to identify articles from four major databases: the ACM Digital Library, IEEE Xplore, Scopus, and Web of Science. We selected the ACM Digital Library and IEEE Xplore because these databases index most major international venues for HCI research, including significant numbers of papers on XR and affective computing. We searched Scopus and Web of Science to capture additional venues that may contain relevant papers (e.g. *Frontiers* journals) and because these databases were used in previous reviews of technologies using biodata [14, 46, 102, 158]. We used our keyword string to capture potentially relevant articles from each database, based on their title and abstract. For Scopus and Web of Science, we focused only on the Computer Science subject area to narrow down the number of returns and limit our scope to exclude research in neighbouring disciplines such as psychophysiology and biology.

## 3.3 Selecting Relevant Articles

Our database searches yielded 4,280 records, which we imported into Covidence[1] for screening. Covidence automatically removed a total of 1,876 duplicates. The full screening process is shown in Figure 2.

In the first phase of screening, the first and third authors independently assessed the relevance of all 2,404 articles by their titles and abstracts. Papers were retained if they satisfied the following inclusion criteria: (1) incorporated any form of biodata, (2) focused on emotions, (3) employed XR technologies, and (4) were published in English. The proportionate agreement between coders at this stage was 0.76. Disagreements arose over the interpretation of XR technologies and whether systems should only be based on immersive head-mounted displays. Such cases were excluded after team discussion, reflecting the sense-making and consensus-building that typically occur early in a scoping review. In the same vein, we excluded papers that did not employ biodata, such as those that measure empathy-arousing by self-reporting [76]. We further excluded those that used physical inputs like gesture [71] or other behavioural measures of physiology, which represent external affect measurements rather than physiological data [117]. This resulted in 1,191 articles for the full-text screening.

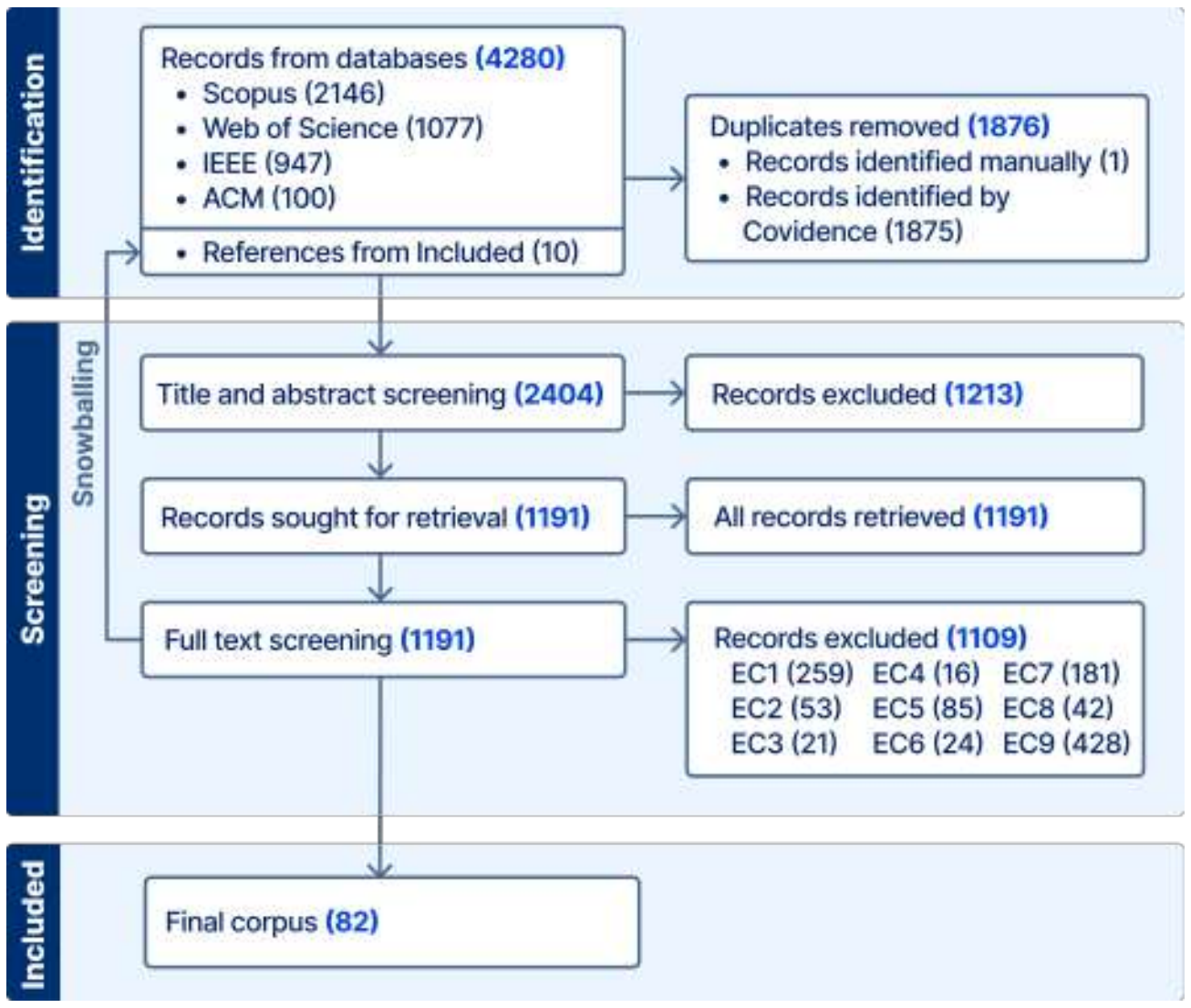

**Figure 2: PRISMA diagram exported from Covidence, illustrating the review process from source identification to the final corpus.**

In the second phase of screening, the same two authors independently evaluated the full text of all remaining 1,191 articles to determine whether they should proceed to data extraction. We defined a total of nine exclusion criteria to identify irrelevant articles (see Appendix B). All authors met regularly to establish and refine the exclusion criteria based on an unfolding mapping of the literature. The proportionate agreement was 0.91 at this stage. Disagreements involved papers that provided datasets without accompanying functional systems [34, 167], or used biodata merely as a study measurement, for example, dependent variables without biofeedback integration [33, 114]. These were excluded and resolved through consensus.

Lastly, we engaged in backward snowballing [4] by examining the reference lists of articles that were being marked as relevant. This is an established technique in scoping reviews and has been widely used by the HCI community [e.g. 140, 160]. From this, we identified an additional 10 articles that appeared highly relevant but were not captured by our search string due to terminology differences. For example, some articles had design intentions to make people aware of their internal state but talked about stress [87, 149] and anxiety [154] as sources of emotions. These articles were subjected to the aforementioned phases of title, abstract, and then full text screening. All 10 articles were included in the final corpus.

[1]Covidence. Systematic review management tool. https://www.covidence.org/

## 3.4 Charting the Data

A total of 82 papers were included in the final corpus[2]. We developed a data extraction template to capture a range of information from the sources and begin answering our research questions. This included data such as publication venue and year, along with information about each paper's aims, methods, and findings. We also captured a descriptive summary of the XR system presented in each paper, its intended purpose, and the technologies used. The data extraction template was created by the first author and was then independently piloted and amended by four other authors on eight papers (two per author). We iteratively revised the template based on team feedback, before importing the final version into Covidence. The first and third author then completed the data extraction template for all 82 papers. The first author also read each paper in full. We exported the consensus dataset from Covidence (in the form of a spreadsheet) for subsequent data analysis. The final version of our data extraction template can be found in Appendix C.

## 3.5 Collating and Summarising the Results

We analysed the charted data through categorisation, quantisation, and thematic analysis. We began by deductively categorising the papers and used summary statistics to highlight key features of the literature, including publication venue, type of biodata used, emotion studied, the frequency of specific XR technology use, and the evaluation approach of each study. We then inductively and collaboratively conducted an interpretive thematic analysis to develop an understanding of goals and challenges identified across the reviewed studies. We iteratively coded and discussed the extracted data, reaching team consensus on four main goals underlying the design of affective XR systems and four key challenges that impact users' experience.

To analyse the system designs and their functionalities, we began by using Figma to create a visual clustering of all system screenshots presented in the source papers. These were organised into separate pages with notes. The first author then studied the data flows within the affective XR systems, having observed a key distinction between single- and multi-user systems while reading the papers. The analysis was subsequently refined and expanded through team discussions. During this process, we found Moge et al.'s [95] existing framework useful to characterise how biodata was collected and shared across the systems. We arrived at an expanded version of this framework to fully classify the ways that affective XR systems can support emotion representation based on biofeedback.

We next analysed design dimensions of systems reported in the literature. We began by clustering all systems based on a distinction between 'literal' versus 'mapped' use of biodata. We then reviewed all visual materials with annotations and team deliberation to identify additional dimensions that represented points of overlap and distinction between the systems. This process continued iteratively until we could not identify any further dimensions, suggesting exhaustion of the design space. During this process, some tentative dimensions were collapsed and discarded. An example discarded dimension was 'Same Device vs Different Devices', which arose because some systems use different technologies [24, 80] while others use identical hardware [28, 130]. However, we excluded this dimension because it was a hardware-driven choice rather than one related to emotion representation. From this analysis, we derived a set of four goals, ten interaction flows, nine design dimensions, and four challenges, which together provide a conceptual mapping of the affective XR literature. We present these findings in the following sections.

# 4 The Landscape of Affective XR

This section presents results based on categories defined by our scoping review template, corresponding to the summary data and information recorded through charting the literature. In subsection 4.1, we provide a descriptive overview of the included literature. We first answer RQ1 by examining the stated goals before discussing user interaction flows and design dimensions of affective XR systems in section 5. Our analyses indicate this growing body of affective XR covers a diverse landscape of biodata and emotions, including a range of systems with notable attention on cardiac measurements, positive-negative affective continua (e.g. relaxation-stress), VR-dominated implementations, and common evaluation methods including SAM and PANAS. Then subsection 5.3 answers RQ2 by highlighting the key challenges.

## 4.1 Characteristics of the Literature

*4.1.1 Historical Progression.* Our review process captured a total of 82 publications. Although we did not impose a date range restriction during our search, the retrieved articles span from 2013 to 2025, with no relevant publications identified in 2014. The papers were published across 43 different venues. The most frequent venues were ACM CHI ($N$=14, 17.1%), IEEE VR ($N$=8, 9.8%), SIGGRAPH ($N$=7, 8.5%), ISMAR ($N$=6, 7.3%), VRST ($N$=3, 3.7%), and CHI PLAY ($N$=3, 3.7%). Other venues ($N$=41, 50%) included Augmented Humans, HCI International, IJHCS, DIS, and TVCG.

Figure 3 illustrates the annual distribution of papers by venue. The figure shows an increase in 2017, aligning with the increased availability of consumer-grade VR/AR devices such as the Oculus Rift (used in 40.6% of the articles) and the HTC Vive (used in 25.6%) around that time. The third most-frequently used headset was the Meta Quest 3 (used in 13.4%), which likely replaced the former two after its release in 2023. Additionally, there is a slight decline in publications between 2019 and 2022. This may be due to the COVID-19 pandemic, which made research on XR challenging due to feasibility issues. Despite this temporary drop, the overall trend shows a steady increase in research outputs with interest from multiple HCI and XR research communities.

*4.1.2 Types of Biodata Used.* A total of 17 different biodata types were used in the papers. Table 1 lists the frequency with which each type was used across all articles. Heart rate (HR) [26, 28, 70] and electroencephalography (EEG) [130, 164, 166] were used most often, appearing 35 times each. These were followed by respiration/breathing rate (RSP) [104, 151, 154], heart rate variability (HRV) [9, 87, 152], and electrodermal activity (EDA) [16, 126, 137].

[2]All papers are included in our references section and are marked with a * flag.

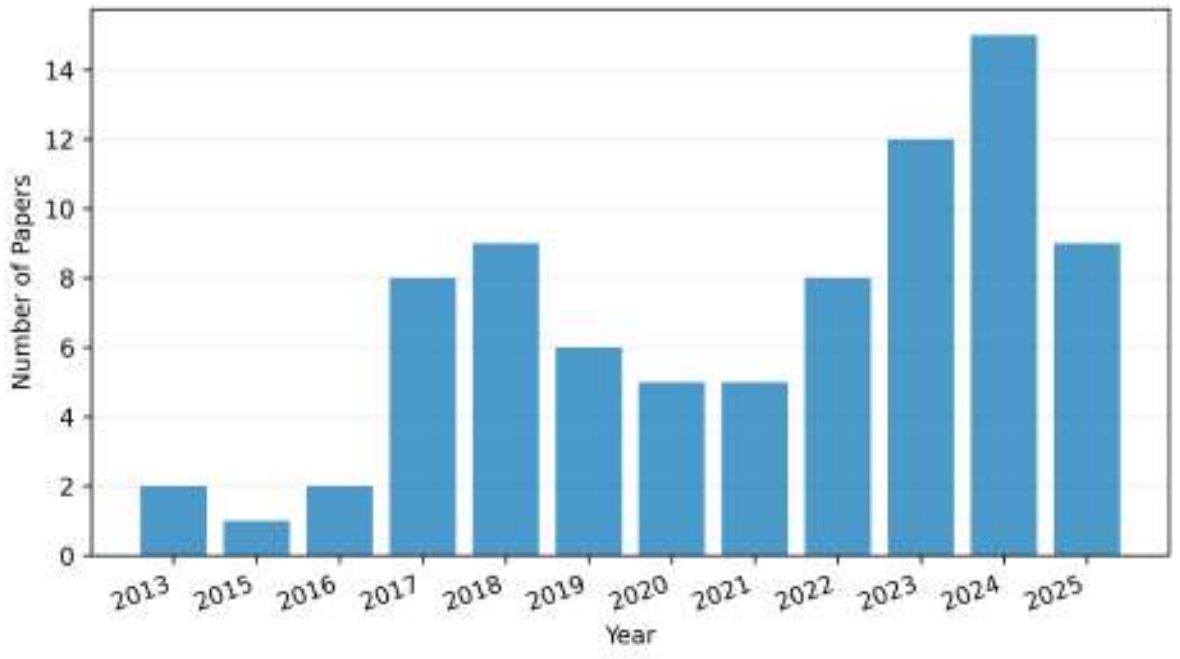

**Figure 3: Number of papers per year by venue (Nb. 2014 is excluded in the figure due to no records in that year).**

We found that 37 papers (45.2%) used a single type of biodata in their system or study. The remaining 45 papers (54.8%) used two or more types of biodata. Examples of combinations included HRV and EDA [16, 126, 137], HR and RSP, [27, 146], and HR, EEG and RSP [30, 120, 150]. Three papers used more than five different types of biodata to represent emotions [10, 37, 85]. For example, Bernal et al. integrated EDA, EEG, electrocardiography (ECG), electrooculography (EOG), and electromyography (EMG), to map users' real-time expression and emotion onto lifelike avatars in a virtual environment [10].

**Table 1: Frequency of single biodata uses in our corpus. Similar biodata are color-coded into the following categories: Cardiac, Neural, Skin, Respiration, Eye, Muscle) (Nb. the total number of instances exceeds 82 because a given article may include multiple types of biodata).**

| Abbr. | Biodata | Counts | Category |
|---|---|---|---|
| HR | Heart rate | 35 | Cardiac |
| EEG | Electroencephalography | 35 | Neural |
| RSP | Respiration/Breathing rate | 19 | Respiration |
| HRV | Heart rate variability | 18 | Cardiac |
| EDA | Electrodermal activity | 18 | Skin |
| EOG | Electrooculography | 13 | Eye |
| GSR | Galvanic skin response | 11 | Skin |
| FE | Facial expression | 6 | Muscle |
| EMG | Electromyography | 3 | Muscle |
| ESR | Eye-blink startle response | 2 | Eye |
| PD | Pupil dilation | 2 | Eye |
| BVP | Blood volume pulse | 2 | Cardiac |
| ECG | Electrocardiography | 2 | Cardiac |
| PPG | Photoplethysmography | 2 | Cardiac |
| BP | Blood pressure | 1 | Cardiac |
| GF | Gaze fixation | 1 | Eye |
| SKT | Skin temperature | 1 | Skin |

*4.1.3 Emotions Studied and Related Biodata.* Figure 4 illustrates the specific emotions targeted for investigation or representation in the included papers, along with the corresponding biodata collected to operationalise these affective constructs[3]. To maintain analytical focus on specific emotional phenomena, instances utilising empathy as a primary research objective were excluded from this analysis [e.g. 65, 83, 89, 124].

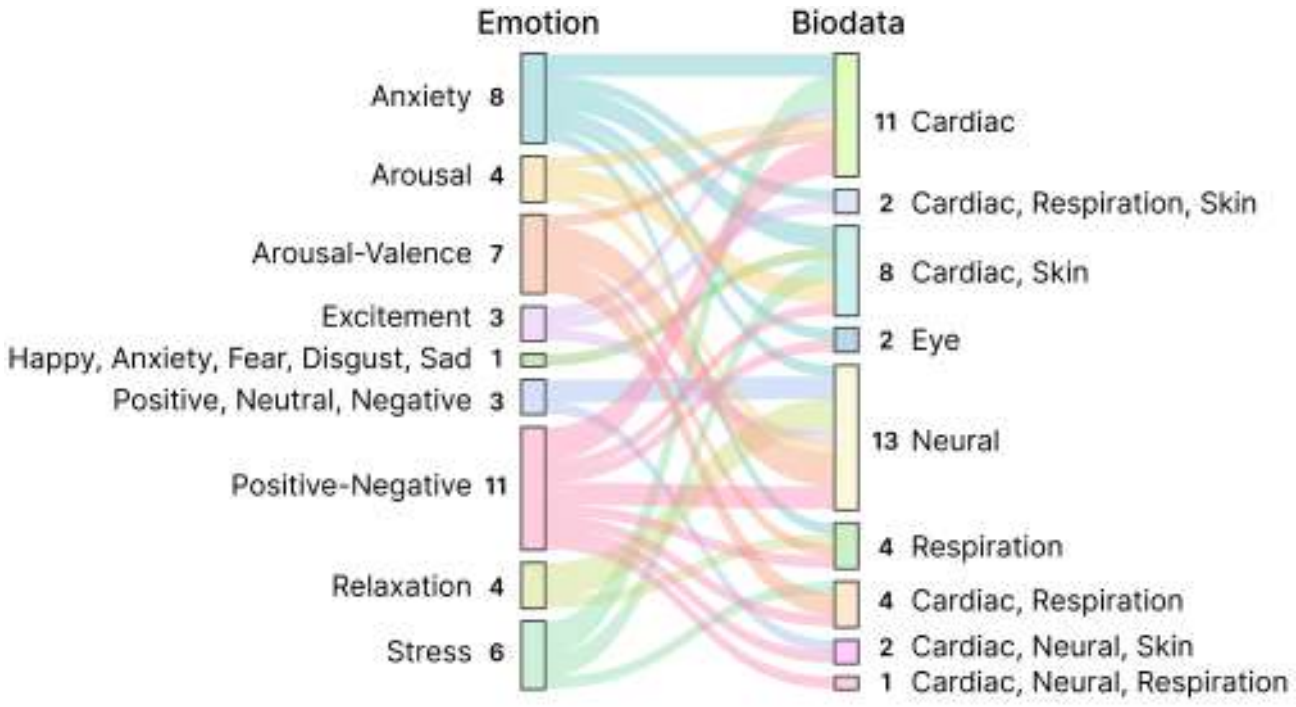

**Figure 4: Sankey Diagram showing the progression and distribution of specific emotions, as labelled in the source paper, and the corresponding biodata.**

This leaves 47 papers for examining the relationship between emotions and biodata, revealing distinct patterns of biodata use across different emotional states. Notably, cardiac measures are predominantly employed in "Positive-Negative" affective continua (e.g. calmness-agitation [45], joy-fear [28], and relaxation-stress [104, 135]), "Anxiety" [35, 111, 154], and "Stress" [87, 152]. Conversely, neural measures are predominantly employed in the "Arousal-Valence" frameworks [60, 82, 130], and "Relaxation" [7, 91]. A notable observation concerns the ubiquitous presence of cardiac measurements across all multimodal biodata studies, suggesting its established role as a foundational physiological marker that researchers consistently incorporate.

*4.1.4 XR Technologies and Application Domains.* To identify technologies used, we classified the papers according to their explicitly stated approaches (i.e., VR, AR, or MR). We observed that one paper utilised XR terminology [23], which we categorised as MR based on the content of the paper and a widely accepted definition of MR [141]. Overall, 61 papers used VR, 11 employed MR, and 7 applied AR. The remaining 3 papers involved combinations of devices, comprising VR and AR [10], VR and a screen-based projection [144], or AR with a laptop [24].

Figure 5 shows the frequency with which particular XR technologies were used within different application domains. The most prevalent application domains were *Gaming and entertainment* (27.4%; e.g. [1, 18, 55, 61, 89, 97]) and *Healthcare and medical* (26.5%; e.g. [59, 82, 111, 135, 150, 154]). These were followed by *Training and education* (20.4%; e.g. [99, 125, 146, 152]) and *Collaboration* (15%; e.g. [27, 52, 65, 166]). We identified a diverse set of *Other* scenarios (10.7%), including social communication [130], interactive art [83], and architectural design [7]. In terms of where XR technologies were commonly applied, VR was most frequently used in gaming and entertainment, AR was used most often for healthcare and medical-focused research, and MR was most applied to collaboration scenarios.

[3]Some of the papers provided their data as on open dataset, which we list in Appendix D along with other datasets screened out at earlier stages of our review.

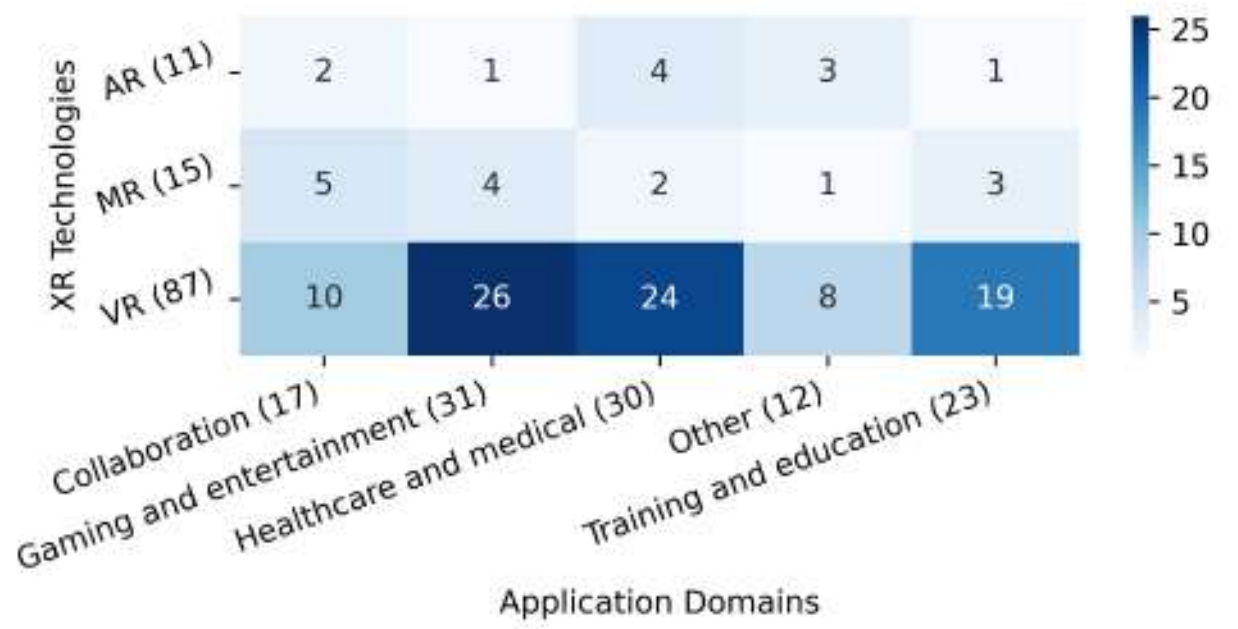

**Figure 5: Heatmap showing application domains and frequency of XR technology use within each domain. The numbers on the X and Y axes show the total number of that type. The total number exceeds 82 as a given publication could be coded as about both collaboration and training (e.g. [80, 125]).**

*4.1.5 Evaluation Methods.* We analysed the methodologies employed in the 82 papers to understand the range of measures used in the studies[4]. Measurements of emotion were the most commonly employed, appearing in a total of 52 papers. Some studies used standardised or established scales including the *Self-Assessment Manikin* (SAM, $N$=15), *Positive and Negative Affect Schedule* (PANAS, $N$=13) or the *State-trait Anxiety Inventory* (STAI, $N$=8).

These questionnaires were applied based on the goals of the study: some were used as baseline measures of emotion to identify changes in affect after exposure to an intervention, whereas others were used solely for evaluation. Scale choices also depended on the number of users involved; some were applied to evaluate self-reported emotions, whereas others were used to gauge the affective experiences of teammates or collaboration partners.

In addition to established scales, 12 papers used a customised questionnaire based on the needs of their user study [7, 97, 133]. Outside of emotion measurement, studies employed other metrics depending on their needs. For example, health-oriented systems used measures of mindfulness [86, 120] or mental health [59], while applications for gaming and teamwork used experience questionnaires [48, 61, 149] or measures of task load [15, 125, 153].

## 4.2 Goals for Affective XR

This section describes four goals that underlie the design of affective XR systems, providing an overview of what researchers have hoped to achieve when creating these technologies. Table 2 lists these goals and provides example studies. Below we highlight the benefits of affective XR systems for achieving these goals, based on the analysis of the literature.

*4.2.1 Awareness and Understanding.* Many of the systems we reviewed were motivated by the fundamental aim of making people more aware of their emotional states. This was exemplified by systems that translate biodata into a representation of emotion and display it back to the user, as a foundation for further reflection on emotion analysis. For instance, the *Body Cosmos* system captures and processes biodata into emotion indicators to enhance awareness of a person’s ‘cosmic identity’ [83]. Similarly, *Stressjam* used colour variations on virtual VR controllers to provide real-time awareness of stress states [87]. The *Life Tree* system employs breathing feedback to strengthen emotional self-awareness by observing the shape of a tree, which changes based on the biodata [104]. In an evaluation of the system, participants claimed that changes in the representation helped them become aware of their own breathing, and felt the design was more powerful than other applications they had used for breathwork [104].

Several systems targeted emotion awareness for specific user groups or contexts. The *Multi-Self* system supports metacognitive monitoring by evaluating designers’ real-time emotional responses during creative work [164]. Conversely, *Tell Me How You Play* examines how arousal state indicators shape interactive experiences in asymmetrical gaming, involving players using laptops and VR head-mounted displays [18]. In addition, several systems explored the use of artificial intelligence (AI) to enhance emotion awareness. In some cases, this was done with an AI agent. For instance, *Sam* is an AI-driven, autonomously-animated virtual agent designed to elicit empathy [85], and *MiRA* is a mixed reality agent that perceives and responds to empathy within shooting games [16]. One study used generative AI to create personalised virtual restorative environments for stress relief [15], while another combined AI-generated visuals and sound to produce adaptive environments that respond dynamically to the user’s physiological state [30].

In some cases, multisensory integration was used to further enhance awareness of emotions. Kody et al. introduced a system that used both visual and audio cues, based on GSR and HR data, to modify the frequency and intensity of an emergency light and a siren, respectively, to maintain participants’ engagement [161]. Furthermore, haptic quality emerged as a critical factor, participants reported that the *“cool and damp texture of the device felt like a real heart”* and that the *“soft and warm texture of the device made me feel very comfortable”* in a system that aimed to modulate anxiety by two polarizing VR content types and tangible representation of a beating heart in the user’s hand [93, 131].

*4.2.2 Regulation and Induction.* Building upon the foundation of emotional awareness, other systems aim to support users in actively regulating, inducing, or manipulating their affective states. Focusing on emotion regulation, systems have been designed to alleviate negative emotions or reduce stress through relaxation. *Transcendental Avatar* promotes relaxation through an abstract, personified avatar by changing their velocity, shape, and colour; participants described the experience as *“relaxing, sometimes similar to meditation”* or noted that it made them *“sleepy and calm”*, while others found it *“exciting and beneficial to relaxation”* [137]. Designed for people with panic disorder, *Drop the Beat* facilitates coping through controlled exposure to anxiety-inducing scenarios [131], and *Emotion Viruses* employs catharsis therapy within a VR serious game to help users alleviate negative emotions [168]. In cases of extreme emotional states, Hart et al. intervene by altering avatar expressions, for instance, representing exaggerated anger with a 'monster face' [51]. Wang et al. developed an EEG-based mindfulness system that guides users toward emotional regulation; one participant remarked, *"The techniques taught helped me calm down during stressful situations"*, while others reported that natural sounds (e.g. insects and birds) contributed to a sense of inner peace *(P14, P15)* [157].

[4]All papers provided their measurements, which we list in Appendix E.

**Table 2: Four underlying goals driving the design of affective XR systems, based on each paper's stated research aims and our interpretation analysis.**

| Goals | Description | Example Studies |
|---|---|---|
| Awareness and Understanding (*N*=20) | Aims to recognise, interpret, and recall emotional states, primarily self-directed or interpersonal. Emphasises cognitive identification and comprehension of emotions, rather than active regulation. | • *Life Tree* [104] visualises growth and the form of leaves based on the user's breathing for relaxation.<br>• *Stressjam* [87] enables users to understand their current HRV state through colours in the virtual controllers, e.g. blue indicates 'calm mode', orange for 'stress mode'. |
| Regulation and Induction (*N*=30) | Aims to modulate, manipulate, or intervene in emotional states to improve psychological well-being or induce context-specific responses. Unlike awareness, it involves actively mitigating negative states or eliciting desired emotions. | • *AR Music Box* [150] guides users' breathing for emotion regulation by its turning and music.<br>• *Drop the Beat* [131] facilitates coping through controlled exposure to anxiety-inducing scenarios by a simulated haptic feedback system. |
| Social Interaction, Communication and Empathy (*N*=12) | Focuses on emotional exchange, shared empathy experiences, and interpersonal connection through emotional cues. It highlights relational and collaborative aspects of emotions within virtual social environments. | • *Empathy Glasses* [80] employs multimodal feedback to enhance emotional connection with partners in remote collaboration.<br>• *Neo-Noumena* [130] augments interpersonal communication by displaying dynamic fractal visuals that change in colour and position according to users' brain activities, thereby fostering empathy. |
| Adaptation and Control (*N*=20) | Refers to systems that dynamically respond to emotional states to adjust task level, challenge difficulty, or environmental parameters. It distinctly employs emotional input as a control mechanism to facilitate progression and interaction in adaptive systems. | • *AmbuRun* [1] controls the game difficulty by the user's neurofeedback to vary the level of excitement.<br>• *HeartFortress* [61] adjusts the game levels continuously based on the user's arousal state, creating an engaging affective experience. |

Conversely, systems use biofeedback to induce emotional states intentionally, such as by evoking happiness from a neutral baseline. Dey et al. examine the effects of manipulated audio heart rate feedback (e.g. increased, decreased, or accurate) across diverse emotional contexts (e.g. happy, anxiety, scary, disgust, sad) [25]. Similarly, *Feels Like Team Spirit* elicits positive shared experiences across asymmetrical player roles and demonstrates the utility of biometric feedback for self-regulation [70].

*4.2.3 Social Interaction, Communication, and Empathy.* A third motivation for affective XR systems was for physiological signals to serve as inputs to systems that aim to enhance social interaction among users. In many cases, this is framed in terms of fostering empathy through understanding another person's emotional state.

Many studies report that sharing biodata improves mutual understanding and communicative engagement. AR technologies, for instance, have been designed to mediate emotional communication during face-to-face interaction. *Neo-Noumena* augments interpersonal communication by displaying dynamic fractal visuals that change in colour and position according to users' emotional states, thereby fostering empathy [130]. In contrast, *AuRea* adopts an asymmetric design in which only one person can perceive their partner's emotional state [153]. Those who saw their partner's current state reported a stronger sense of connection during a collaborative task, and the system was shown to increase cognitive empathy.

In shared virtual environments, systems such as *DYNECOM* use bio-adaptive visualisations to evoke compassionate and empathic feelings toward avatars representing partners, suggesting a form of affective contagion through mediated feedback [124]. Similarly, *Empathy Glasses* support remote collaboration by sharing gaze, facial expressions, and physiological signals between users [80]. In an evaluation of the system, one participant noted, *"I could easily point to communicate, and when I needed it, I could check the facial expression to make sure I was being understood."* [80]. *My Heart Will Go On* enhances social connectedness by visualising heart rates in a collaborative escape game, reinforcing a sense of shared experience [55]. Salminen et al. further demonstrate that shared physiological data in social VR can strengthen empathy and behavioural synchrony among users [123].

Auditory and haptic feedback also play significant roles in shaping socio-emotional experiences. Dey et al. combined these modalities in a system that enabled people to hear and see their collaborator's heart rate [26, 27]. In an evaluation, one participant expressed, *"... it is great to feel my collaborator's heart rate ... makes me feel I am not alone!"* [27], though another commented, *"... it is interesting to know the other person's heart rate, but I do not think it changed my actions at all in the given circumstances."* [26]. An emerging approach involves the use of historical biodata to cultivate empathy and emotional recall. *Re-Touch* integrates VR, haptic feedback, and biodata to enhance memory recollection [47].

*4.2.4 Adaptation and Control.* The fourth motivation involved transforming biodata into novel control modalities, enabling dynamic adaptation of gameplay, virtual environments, and interaction experiences. These systems utilise emotional and arousal states as implicit input to adjust various aspects of the system in real time, often without requiring explicit user commands.

Some examples employ real-time biometric feedback to dynamically adapt both gameplay difficulty and environmental parameters. In *HeartFortress*, challenge levels are modulated continuously based on the player's arousal state: during periods of elevated HR, enemy movement speed is gradually reduced via a logistic mapping function to support player re-engagement [61]. Conversely, when EDA signals indicate declining arousal, the system increases both the spawn rate and dispersal range of enemies to maintain engagement. In an evaluation of the system, 90% of their participants reported perceiving adaptive changes in enemy quantity, while 70% noted variations in movement speed, reflecting not only user awareness of the emotional adaptation mechanisms but also an overall appreciation for the responsive and dynamic gameplay experience [61]. Similarly, *AmbuRun* employs neurofeedback to influence game speed and difficulty [1], and *WizardOfVR* modifies environmental features such as fog density and visual effects according to real-time emotional input [49]. These approaches illustrate how affective signals can be used to tailor interactive experiences responsively and autonomously.

Beyond serving as a basis for difficulty adaptation, biodata has also been utilised as a direct control mechanism in affective systems. In the *STRATA* system, users regulate their breathing to achieve calmness, which directly enables virtual levitation and progression through the environment [32]. Similarly, *BC-invisibility Power* employs direct brain–computer interaction to control optical camouflage [91], while *Alice Beyond Reality* uses heart rate as an input mechanism, allowing young adults to learn emotion regulation by controlling their avatar's size in a VR game [45]. Similar systems have been explored in collaborative and educational settings. *PhysioHMD*, for instance, maps real-time facial expressions and emotional states onto social VR avatars to enhance non-verbal expressivity and communication [10]. Similarly, *Emotion Superpower* amplifies environmental feedback in collaborative VR tasks by using facial expressions as a form of consensus input among multiple users [67].

# 5 Design Dimensions and Challenges in Affective XR

In this section, we present an analysis of biofeedback designs used in affective XR systems, structured according to two key parameters. First, we discuss the differences in the strategies for sharing emotions evident in the literature. We do this by extending the existing framework of Moge et al. [95] to show how biodata can be differentially shared between users in affective XR. Second, we highlight dimensions that characterise existing biofeedback representations, enabling us to explore points of distinction and design opportunities in our discussion section.

## 5.1 Intra- and Interpersonal Biodata Sharing in XR Systems

This section shows how the flow of biodata in affective XR systems has been organised to create a range of biofeedback mechanisms. Our analysis is based on a categorisation of the 82 papers in terms of whether the system was designed or presented for single-user or multi-user, and how the biodata was shared between the people involved. We further analysed the multi-user designs and identified that biodata can be shared either symmetrically, where users see identical information, or asymmetrically, where users see different information from each other.

To make sense of these distinctions, we present an expanded taxonomy based on Moge et al.'s framework [95], which identified five different approaches used in social biofeedback systems. Our taxonomy expands this to 10 different approaches used in affective XR. Figure 6 shows our taxonomy, based on a similar illustration used by Moge et al. [95]. We also provide the schematic representations of all system screenshots from the final corpus sorted by these categories in Appendix F.

*5.1.1 Single User.* This category represents the most straightforward type of biofeedback system, wherein biodata is collected from a single user and presented back to the same individual, forming a closed-loop feedback mechanism. A total of 53 systems (64.6%) in our corpus involved a single-user design.

① **I feel mine**. Systems in this category enable users to perceive their own affective states through real-time biofeedback. For example, *Stressjam* shows players their own HRV data, and requires them to modulate their stress levels intentionally to progress through the game [87]. Similarly, Qing et al. used EEG signals to quantify emotional states during VR meditation exercises, with adaptive feedback generated to support emotional regulation by showing users their own EEG data [157]. Other systems integrate multiple biosignals, such as HRV, EDA, and RSP, to adapt virtual environments dynamically [111]. This continuous feedback allows users to learn self-regulation strategies, reducing anxiety and enhancing immersion through real-time performance adjustment.

*5.1.2 Multi-user Asymmetry.* This category extends single-user systems by incorporating multiple users, wherein one user's biodata is captured and shared with at least one other person. In our corpus, 17 systems (20.7%) fit this category, corresponding to sharing types ②–⑤ in Figure 6.

② **I feel yours**. These systems capture biodata from one user and present it to another, often to support empathy or cooperative tasks. For example, *Empathic AuRea* takes one person's ECG data and presents it to a primary user [153]. The ECG data is visualised as coloured circles around the interlocutor's head, and is intended to support understanding of emotions during face-to-face collaboration. In another example involving a virtual *Bomb Defusal* game, one player (an instructor using a laptop) shares their facial expressions via an avatar with a VR user, who must defuse a bomb while perceiving the instructor's emotional state [51]. Other systems leverage historical biodata for shared emotional experiences. For example, VR music applications visualise multiple biodata (e.g. BVP and EDA) from audiences in a recorded concert, allowing a single VR user to perceive the collective emotional responses of previous attendees [89, 90].

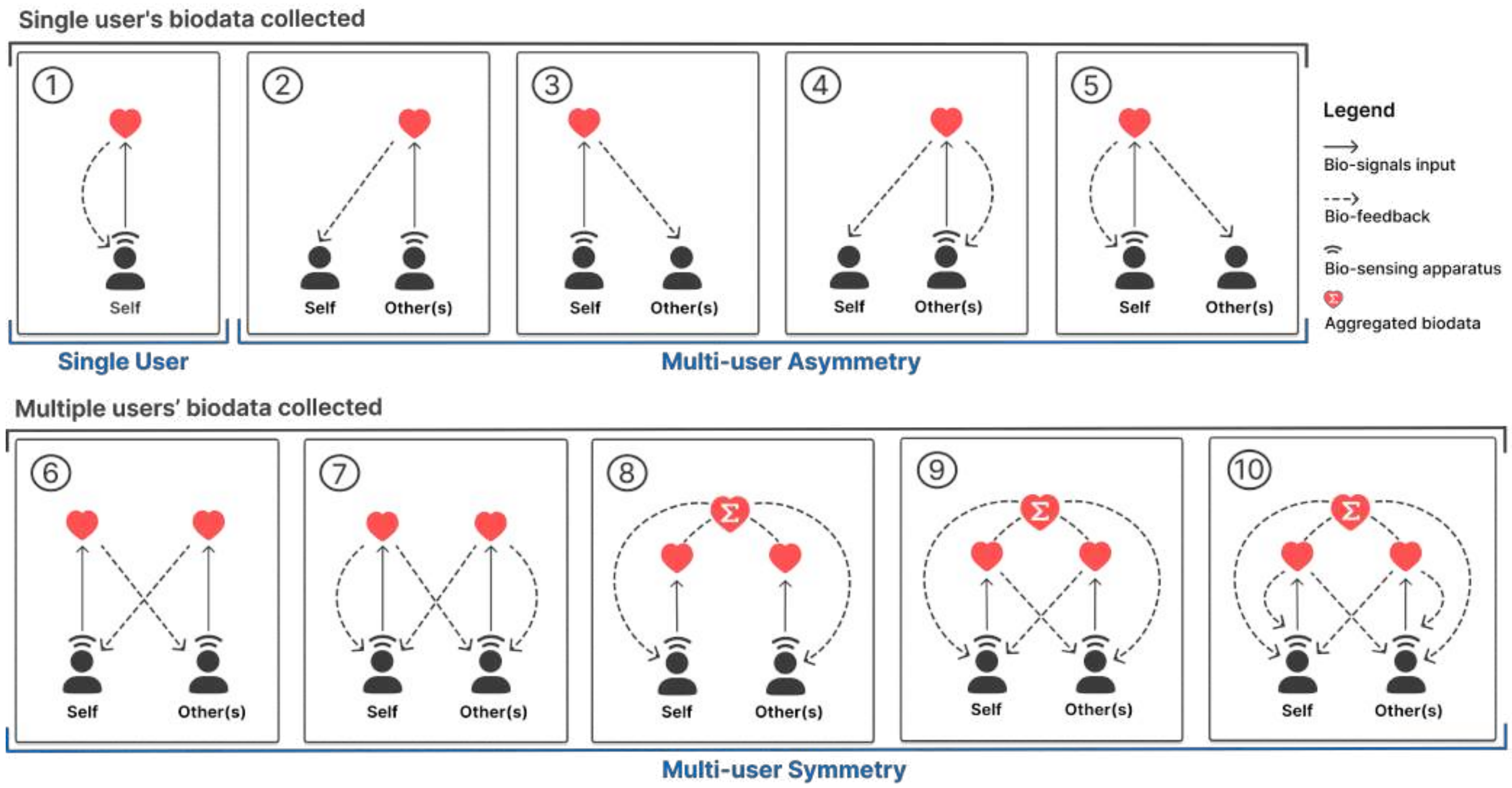

**Figure 6: Ways of biofeedback sharing among users. Giving a heart icon as an example of biodata. *Single User*: ① I feel mine. *Multi-user Asymmetry*: ② I feel yours; ③ You feel mine; ④ We feel yours; ⑤ We feel mine. *Multi-user Symmetry*: ⑥ I feel yours, you feel mine; ⑦ We feel each other; ⑧ We feel ours aggregated; ⑨ I feel yours, you feel mine, and we feel ours aggregated; ⑩ We feel each other, and ours aggregated.**

③ **You feel mine**. Systems in this category capture a user's biodata and share it with others, often to support monitoring or collaborative adaptation. This setup was used most commonly in health settings involving a therapist and a patient. For example, a VR-based post-stroke rehabilitation system integrated EEG to analyse patient emotions (e.g. frustration, depression, relaxation), enabling remote therapists to adjust their behaviour based on inferences about a patient's physical and emotional state [133]. Similarly, a *Truck Driving Simulator* for PTSD therapy uses real-time heart rate to dynamically adjust exposure intensity, allowing clinicians to modulate difficulty based on the patient's stress level during prolonged exposure therapy [69].

④ **We feel yours**. Extending category ②, this approach involves sharing others' biodata with both themselves and a primary user. Only one paper in the corpus adopted this approach. In that study, Dey et al. visualised one VR player's heart rate for an observer in two collaborative games: *Butterfly Catching* (calm mode) and *Zombie Shooting* (stressful mode) [28]. The observer could see and hear the HR feedback of the primary player, while the player did not receive biofeedback from the observer. This design enhanced the observer's sense of connection and understanding of the player's emotional state, particularly in scary scenarios where heart rate visualisation improved empathy and situational awareness.

⑤ **We feel mine**. Extending category ③, this approach involves sharing one's biodata with both oneself and others. This approach was used in only one paper, the aforementioned study by Dey et al., where the system also enabled a player to share their heart rate data with an observer, allowing both individuals to perceive the player's physiological state [28]. While the player receives real-time audiovisual feedback of their own heart rate, they remain unaware of any biofeedback from the observer. Like category ④, this approach remains relatively unexplored, with no additional instances identified within our corpus.

*5.1.3 Multi-user Symmetry.* These systems extend multi-user interaction by capturing biodata from multiple users and enabling mutual or structured sharing among them. A total of 12 systems (14.7%) fit this category, with one also involving an asymmetrical setup. This category corresponds to sharing types ⑥-⑩ in Figure 6.

⑥ **I feel yours, you feel mine**. This category involves capturing biodata from two users and sharing it bidirectionally, but without users being able to see their own data. For example, *CardioCommXR* captured and shared HRV between two VR users performing a collaborative memory-matching task, strengthening interpersonal communication through mutual stress visibility [152]. In another example, Hart et al. tested mutual biodata sharing in a mixed reality car repair scenario, where an expert using an AR headset and a client using a mobile phone with AR effect to see each other's facial expressions, facilitating remote assistance through mutual emotional awareness [52].

⑦ **We feel each other**. This category enables users to perceive both their own and their partner's biodata simultaneously. For instance, *Tell Me How You Play* visually conveys real-time arousal states, derived from EEG signals, for both PC and VR players, allowing each to see their own and their partner's biofeedback across different devices [18]. This design was found to foster social connection and co-presence. Similarly, *My Heart Will Go On* embeds heart rate visualisation spatially and temporally within a VR escape room game, providing implicit navigation cues and supporting non-verbal social understanding among asynchronous players [55].

⑧ **We feel ours aggregated**. Systems in this category capture biodata from multiple users, process it into an aggregated form, and share it with all users. For instance, the *Multiplayer VR Live Concert* system enables multiple users to contribute their individual EEG signals in real time, generating shared visual effects that vary individually [97]. Their aggregated feedback creates a collective visual experience, fostering a strong 'sense of unity' among attendees.

⑨ **I feel yours, you feel mine, and we feel ours aggregated**. This category combines bidirectional biodata sharing with aggregated feedback, as seen in approaches ⑥ and ⑧, enabling users to perceive both individual and collective physiological states. For example, *DYNECOM* visualises each user's EEG signals as an aura around their avatar while also representing RSP data synchronisation between partners as a dynamic 'connecting bridge' [123]. This dual-layer visualisation supported empathic feelings towards the partner through multi-cue biofeedback. Such systems applied multiple visual cues in different physiological states, allowing the user to evoke empathic, warm, and compassionate feelings directed to the avatar statue that represented their partners [124].

⑩ **We feel each other, and ours aggregated**. The final category integrates mutual and aggregated biodata sharing, as seen in approaches ⑦ and ⑧, enabling users to access their own biodata, their partner's, and a collective representation. We found two systems that use this approach. First, DiPaola et al. developed a system where both the counsellor (on a large screen with an EEG headband) and the patient (in VR HMD) aim to achieve relaxed breathing simultaneously [30]. Rather than just remotely suggesting how the patient can relieve stress but not showing the counsellor's real-time biodata, here, the counsellor must keep themselves in a relaxed state aligned with their patient, creating a shared bio-responsive experience [30]. Second, the *Jel* system enables two collaborators (one wearing an EEG headband, another using a VR HMD) to view their own, their partner's, and an aggregated RSP feedback visualisation, fostering mutual relaxation and connection through synchronised breathing [144].

## 5.2 Design Dimensions for Emotion Representations

This section focuses our analysis on the designs employed to achieve biofeedback. We present a set of nine design dimensions that we developed based on the systems presented in the literature. These dimensions characterise the emotion representations according to parameters that are under the designer's control, and which reveal a range of opportunities for designing affective XR systems. Figure 7 presents all the design dimensions, illustrating how each one differs. We also provide a complete breakdown of interaction modalities used in Appendix G.

*5.2.1 Visualisation: Literal vs. Mapped.* (See examples in Figure 8). Our first dimension characterises the way biodata is translated into a representation experienced by the user. In literal representations ($N$=26), emotions are conveyed through elements that represent the biodata in a literal manner; for example, by representing heart rate using an image of a beating heart, or representing brain waves using a time domain graph. Conversely, mapped representations ($N$=56) transform biodata into abstract visual designs that require learned interpretation.

Literal designs have been achieved in a number of ways. Taking heart rate data as an example, some systems simply take the numerical reading from a biosensor and display it back to the user, showing 'BPM: 77' [125] or 'HR: 99' [24, 60, 80, 131]. Other systems use text, e.g. with the words 'High heart rate' [107]; symbols like a pulsing heart in 2D [17, 80] or 3D [93, 131]; or a variety of progress bars, with some using different colours like red or green to represent thresholds [65, 73, 110, 152]. Other techniques include emoji to represent feelings [80], facial expressions that are captured and applied to users' avatars [51, 52, 85], and coordinate scales, including 2D map with sections labelled 'Arousal' and 'Valence' [133, 164], or 3D map showing 'Positive, Neutral, Negative' [81].

Conversely, mapped representations involve taking the biodata and translating it into a representation that does not reflect the original source, but which still aims to convey meaning. Examples include mapping emotions to colours, such as by changing the colour of a virtual controller indicator to blue for calm and orange for stress [87]. Other papers map emotions onto specific phenomena in the virtual environment. Examples include using rain to convey sadness [59]; ocean surface calmness to reflect respiration [113, 120]; a virtual music box to pace breathing [150]; the growth of plants in response to rhythmic breathing or relaxation [104, 157]; and light intensity based on breath input [86]. Representations can also be mapped to graphical forms, including emotion-matched images from databases [37], AI-generated exciting visuals [15], dynamic fractal shapes for anxiety-to-relaxation transitions [35] or high/low arousal and valence [130], and multi-user star and petal effects in concerts [97].

*5.2.2 Agency: Passive vs. Proactive.* (See examples in Figure 9). Passive representations ($N$=80) present biodata without providing explicit guidance or instructions for the users, whereas proactive representations ($N$=2) deliver actionable suggestions or directives based on the displayed biodata to guide subsequent user behaviour.

An example of a passive system is *Deep*, a virtual reality game that uses diaphragmatic breathing to control gameplay, designed to help children at risk for anxiety develop self-regulation skills [154]. The system provides visual feedback through an expanding and contracting circle that reflects breathing patterns, but offers no direct guidance during the interaction. In contrast, several systems incorporate proactive elements that instruct users on appropriate responses. One AR application displays a virtual music box that rotates to guide users through structured deep-breathing exercises for stress reduction [150]. When negative emotions are detected, the system explicitly suggests taking a break and provides encouraging instructions such as 'Keep Breathing'. Beyond visual guidance, some systems employ auditory instructions, such as *CAEVR*, which features an orb-shaped virtual companion that provides natural language navigation assistance (e.g. 'walk towards left') during VR photo-taking tasks [48].

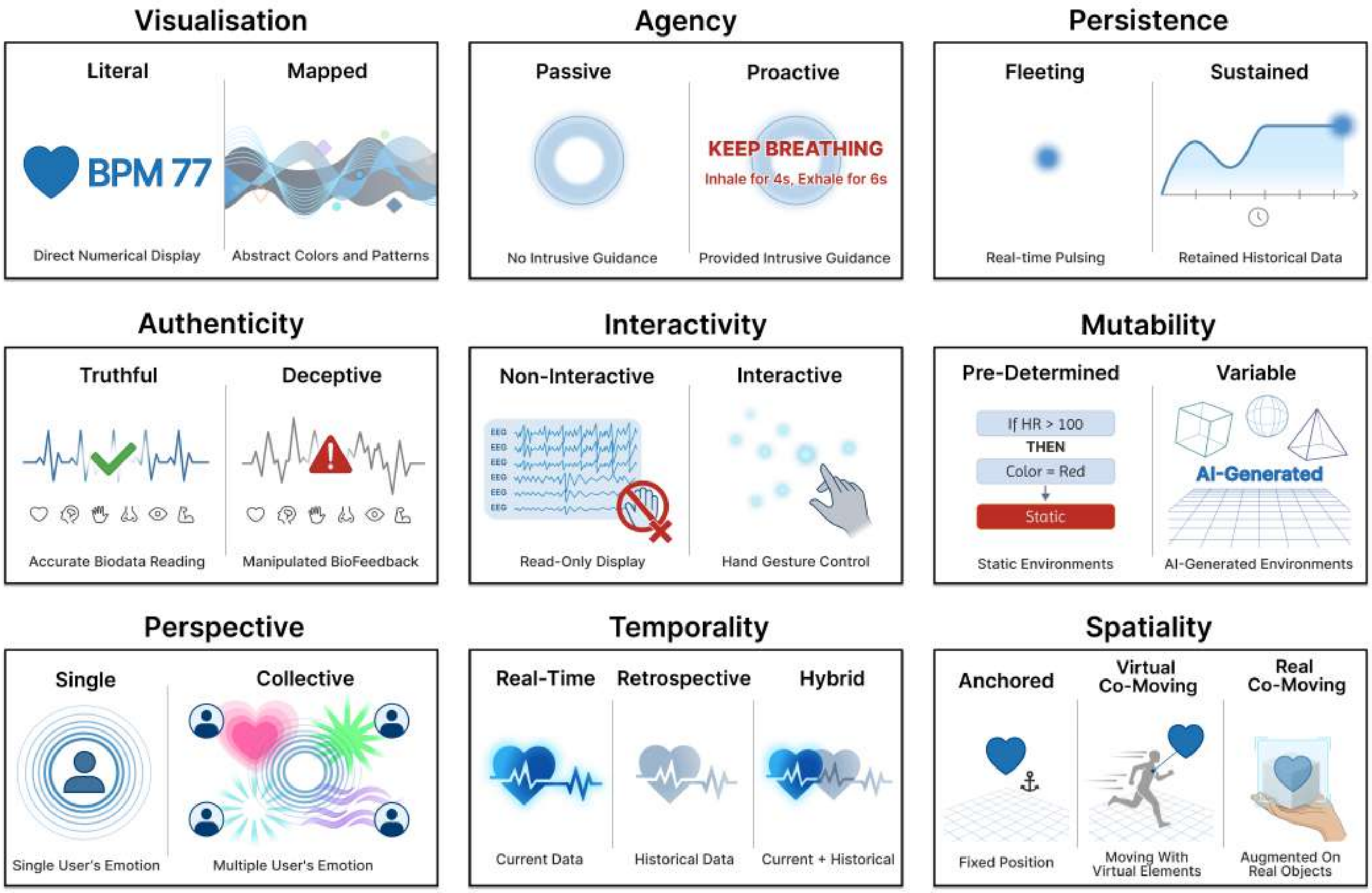

**Figure 7: The set of nine design dimensions for emotion representations. Each dimension comprises a subset of distinct ways to display and convey emotions, with illustrations and a succinct textual description.**

*5.2.3 Persistence: Fleeting vs. Sustained.* (See examples in Figure 10). This dimension describes the difference between designs that show temporary or fleeting representations ($N$=78) of real-time biodata, versus those that create sustained representations ($N$=4) that remain in place over time. The former approach focuses on supporting understanding in-the-moment, whereas the latter retains and visualises historical biodata over extended periods, enabling longitudinal reflection and analysis.

Fleeting designs capture and display biodata that are transiently transmitted, for example, using dynamic visual elements such as pulsating heart icons [17, 24, 28, 93] or a real-time pulsing dot in the quadrant [81, 133, 164] indicating the user's current heart rate data. In contrast, fewer systems involve sustained representations. One example is that of *My Heart Will Go On*, which stores and visualises complete historical heart rate records, allowing users to review their biodata collectively over time [55], while other systems use a simplifier moving line [86, 111]. This approach may help users to understand the trends beyond momentary physiological states.

*5.2.4 Authenticity: Truthful vs. Deceptive.* (See examples in Figure 11). Truthful representations ($N$=77) present biodata accurately as it was captured, preserving the true physiological state of the user (e.g. [60, 80, 125]). In contrast, deceptive representations ($N$=5) intentionally alter or fabricate biodata, displaying artificial or manipulated information for experimental or experiential purposes.

For instance, some deceptive systems have manipulated heart rate feedback by increasing or decreasing the displayed rate by 15–30% to examine its effect on the perception of emotional states during collaboration [25, 26]. Another system enables users to exaggerate emotional expressions, such as transforming their avatar into a monster when anger is detected, to explore how augmented emotional displays impact social interaction and collaboration in VR games [51]. One system incorporated a virtual agent that provides intentionally incorrect feedback (e.g. by falsely indicating increasing skin conductance levels), to examine how deceptive physiological cues influence empathy, social connectedness, presence, and trust [16].

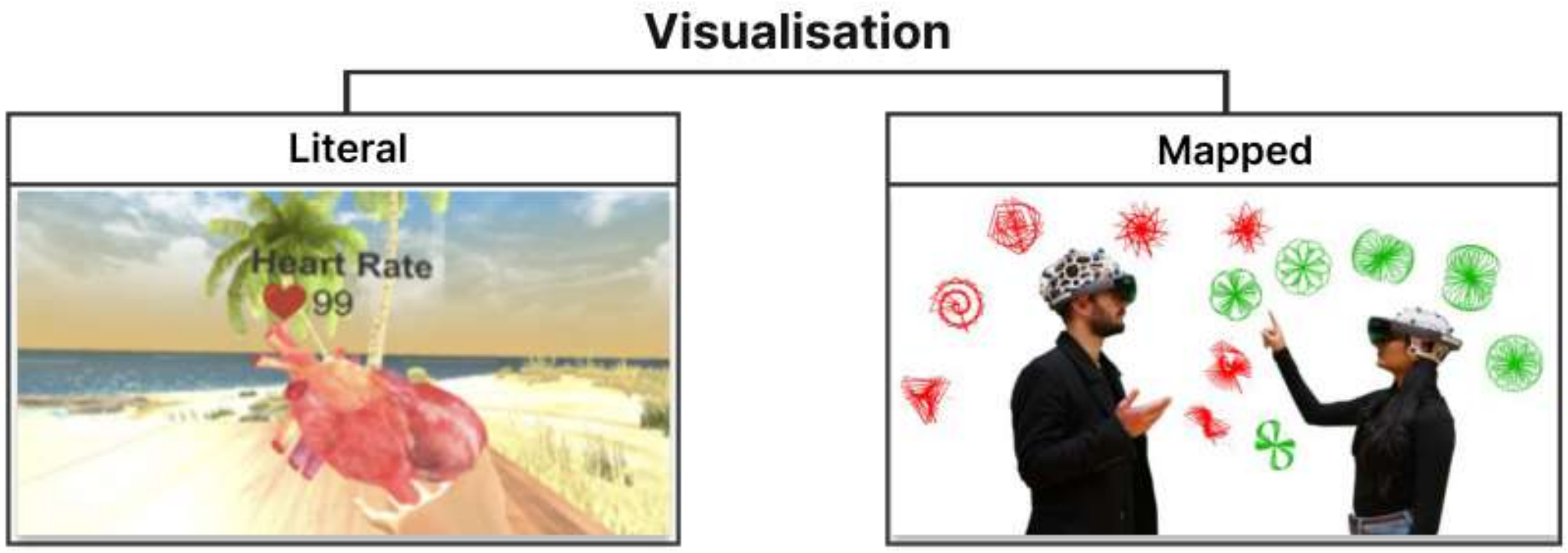

**Figure 8: Examples of affective XR systems illustrating the Visualisation design dimension. Literal: *Drop the Beat* [131] displays straightforward information of a literal message 'Heart Rate', a heart icon, 99 beats per minute, and an anatomically styled heart graphic. Mapped: *Neo-Noumena* [130] maps the user's brain waves onto dynamic fractal shapes with different colours to represent individual affective states in high/low arousal and valence.**

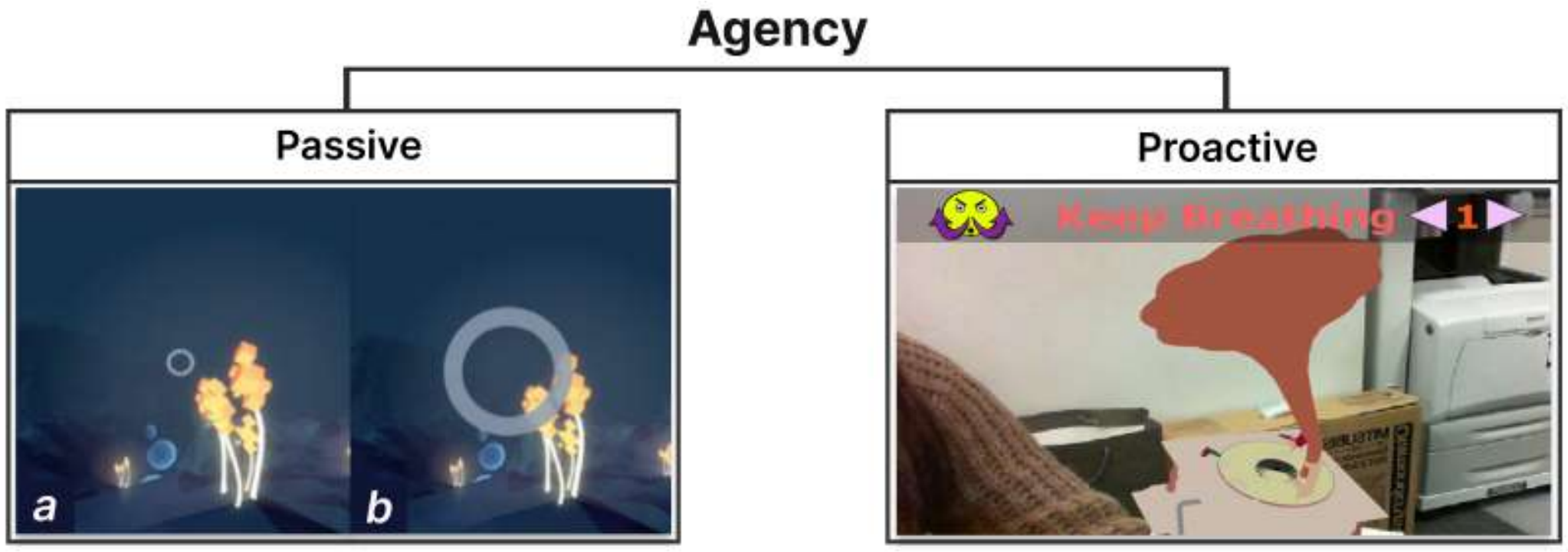

**Figure 9: Examples of affective XR systems illustrating the Agency design dimension. Passive: *DEEP* [154] uses a circle that pulsates to simulate the breathing cycle - contracts for inhale (a), expands for exhale (b) - without explicit instruction for anxiety regulation. Proactive: *AR Music Box* [150] uses a dynamic brown particle on top of a virtual music box, which scales in size to represent breath flow with the explicit textual instruction 'Keep Breathing' for relaxation.**

*5.2.5 Interactivity: Non-interactive vs. Interactive.* (See examples in Figure 12). Emotion representations can be distinguished based on a user's ability to interact with the biodata that is conveyed. Non-interactive representations ($N$=81) present biodata through static or non-responsive visualisations that users cannot manipulate. Examples include designs that present ECG, EEG, or other signals without the ability to interact with those virtual elements physically [1, 45, 61, 69, 80, 104].

In contrast, interactive representation ($N$=1) enables users to engage with and influence their biodata visualisations through explicit physical actions or controls. *BRieFLY* combines guided mindful breathing with interactive virtual fireflies designed to foster self-connection [151]. Users can interact with the visualisation by using hand gestures to push away fireflies or guide their movement, creating a direct embodied interaction with the biofeedback representation.

*5.2.6 Mutability: Pre-determined vs. Variable.* (See examples in Figure 13). Pre-determined representations ($N$=77) employ fixed techniques to display biodata, whereas variable representations ($N$=5) dynamically adapt visualisations based on real-time user states or contextual factors, offering personalised feedback for each moment of interaction. In pre-determined systems, specific emotional or physiological triggers consistently produce the same visual outcome. Examples include displaying larger numerical values for elevated heart rates [24, 125] or assigning specific colours to corresponding emotional states [48, 87, 110, 153].

In variable representations, biofeedback is adapted based on data input or algorithms that control the design of the environment. For instance, one system continuously employs AI to create novel, dynamically generated VR imagery [15]. Such systems shift user attention through context-sensitive content, producing unique visual experiences and continually evolving content tailored to the user's biodata. However, another system utilises AI to convert users' verbal descriptions of memories into personalised visual and auditory content, enabling dynamic revisiting of the memory recall process [47]. This approach leverages retrospective personal narratives rather than real-time biodata, demonstrating how variable

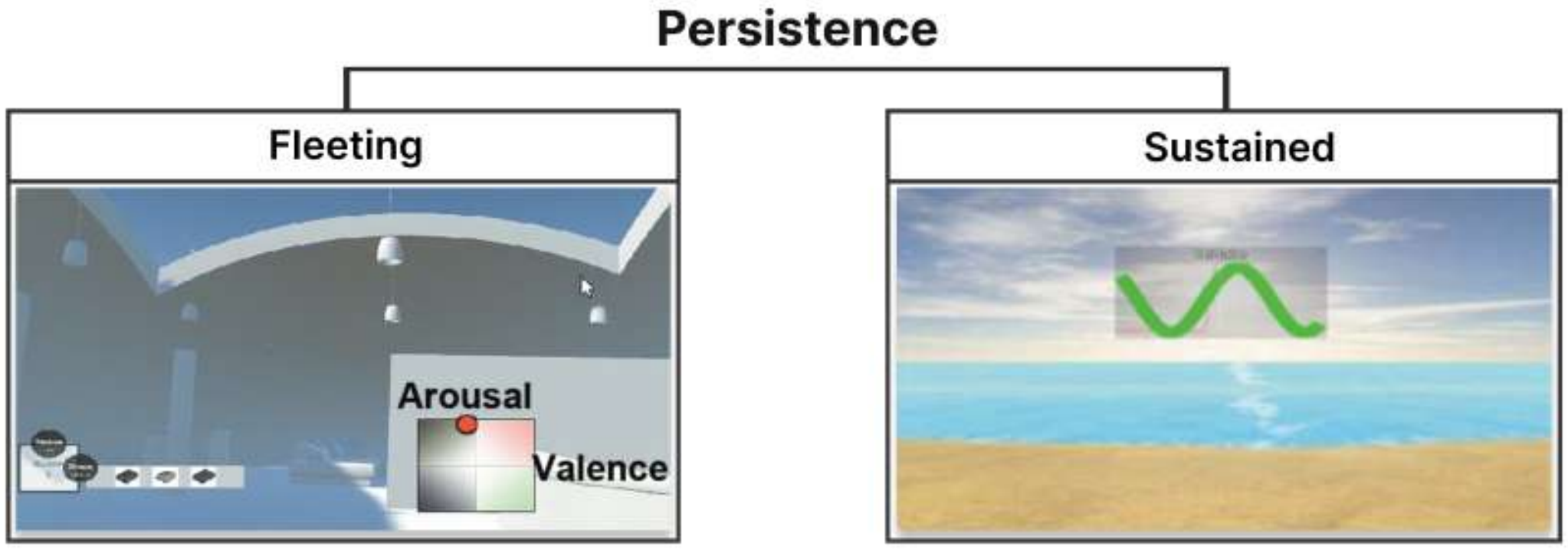

**Figure 10: Examples of affective XR systems illustrating the Persistence design dimension. Fleeting: *Co-Design with Myself* [164] uses a red dot that is real-time pulsing in a quadrant, indicating high/low arousal and valence for reflecting whether the user likes the architectural design, measured by EEG. Sustained: Prabhu et al.'s system [111] uses a moving green line that is provided to show real-time and retained HRV feedback for training to reduce the user's perception of anxiety before and after their surgery.**

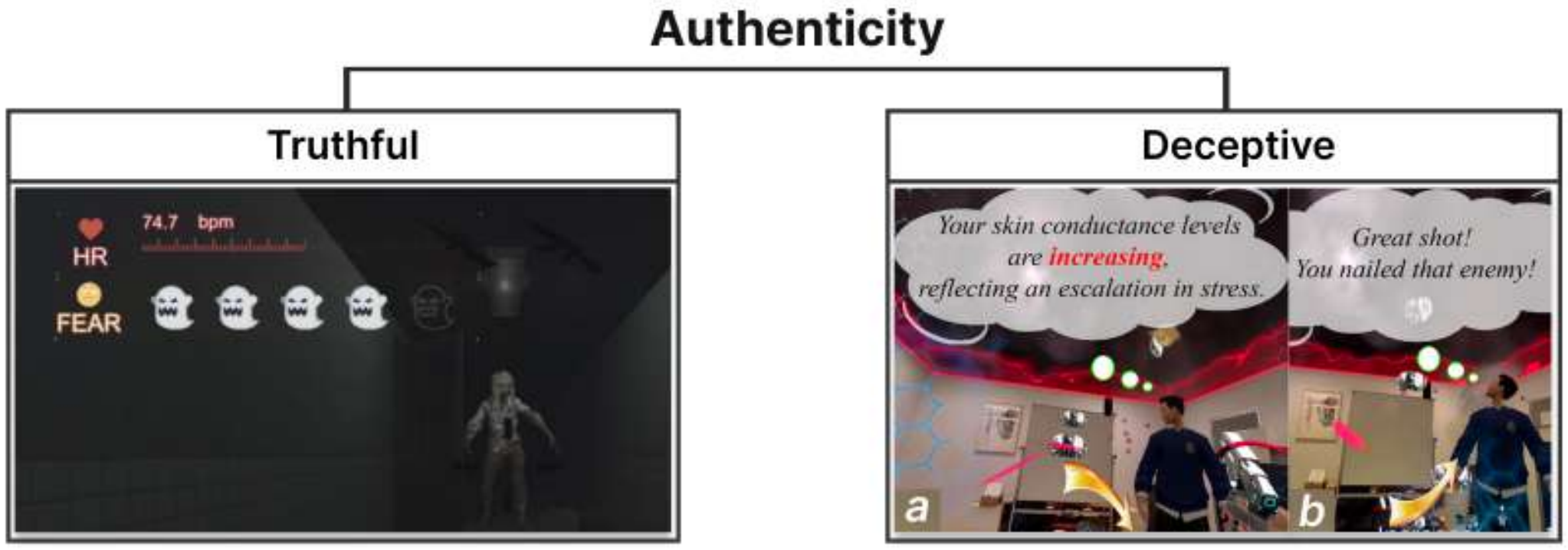

**Figure 11: Examples of affective XR systems illustrating the Authenticity design dimension. Truthful: Inazawa et al.'s system [60] displays real-time heart rate data and fear level, presented without alteration of the biodata. Deceptive: In scenario (a), Chang et al.'s virtual agent [16] perceives empathy in a shooting game by falsely reporting an increase in skin conductance level, while the bottom arrow shows the authentic decrease. Scenario (b) is showing authentic data.**

representations can operate across different temporal dimensions of a user's biodata.

*5.2.7 Perspective: Single vs. Collective.* (See examples in Figure 14). This dimension highlights the difference between systems that visualise biodata from individual or multiple users as distinct elements, or whether the system combines a range of data points into a collective whole. Most systems in our corpus employ single-user representations ($N$=79), displaying individualised biodata through isolated visual elements such as progress bars [73, 110], line graphs [86, 148, 155], or abstract patterns [47]. Even in multi-user contexts, these systems typically present each person's biodata through separate avatars or distinct indicators that remain visually partitioned within the virtual space [10, 11, 123, 124, 137].

In contrast, collective representations ($N$=3) merge multiple users' biodata to generate group-level visualisations. One multi-user VR concert system aggregates audience brain responses to produce unified visual effects (e.g. coordinated stars, heart patterns, or cherry blossom petals) where individual contributions blend into a single experience [97]. This approach transforms separate data points into a new emotional representation that fosters interconnectivity and a tangible sense of unity among users.

*5.2.8 Temporality: Real-time vs. Hybrid vs. Retrospective.* (See examples in Figure 15). This dimension characterises the temporal property of the data shown to users. Real-time representations ($N$=75) present biodata instantaneously, as it is acquired. Retrospective designs ($N$=3) provide a view of historical or recorded data. Hybrid systems ($N$=4) integrate both real-time and retrospective data within a unified design.

Real-time representations straightforwardly capture biodata and display it to the user, such as by showing heart rate as a beating heart icon [17, 24, 28, 93]. Retrospective, in contrast, utilises previously recorded data. For instance, Xiaru et al. captured audiences' emotions during a music concert and created a representation to convey the emotional atmosphere to later users [89, 90]. Retrospective approaches also include applications for individuals, such as

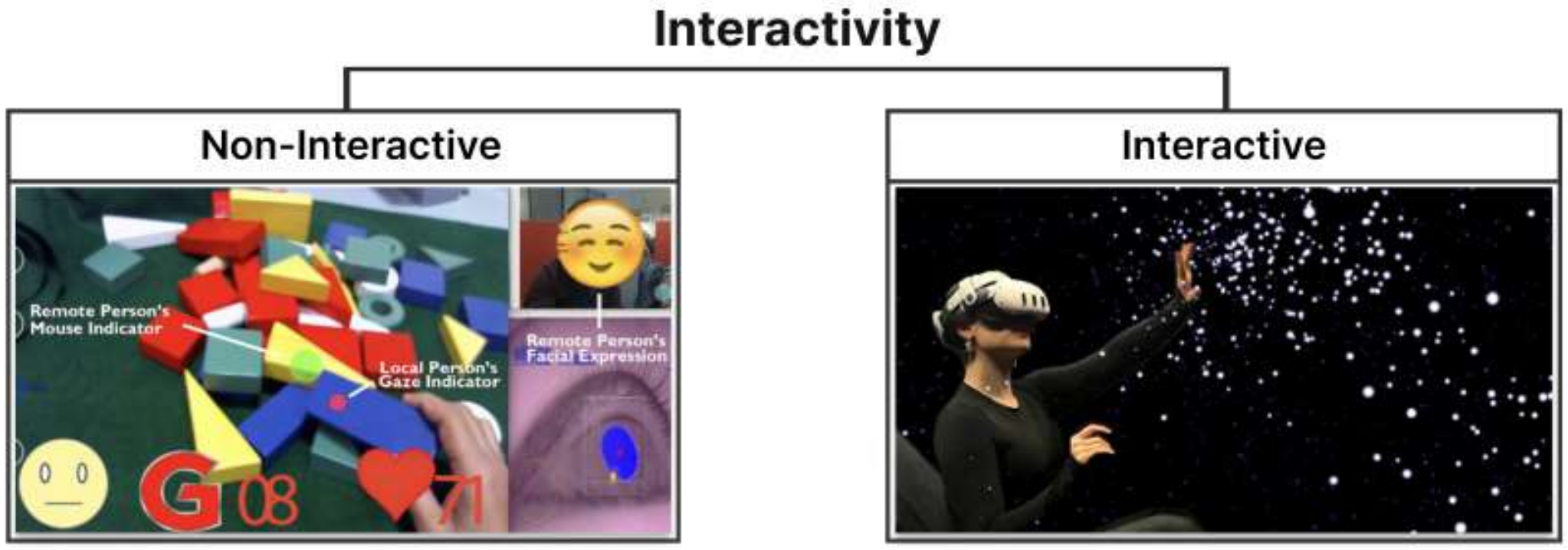

**Figure 12: Examples of affective XR systems illustrating the Interactivity design dimension. Non-Interactive: *Empathy Glasses* [80] directly displays information, including emotional state (e.g. emoji), GSR, and HR in a non-interactive way. Interactive: *BRieFLY* [151] uses hand tracking to enable users to interact directly with a swarm of fireflies through breathing for self-relaxation. As the flock moves by hand, it transitions from purple to white, then back to purple.**

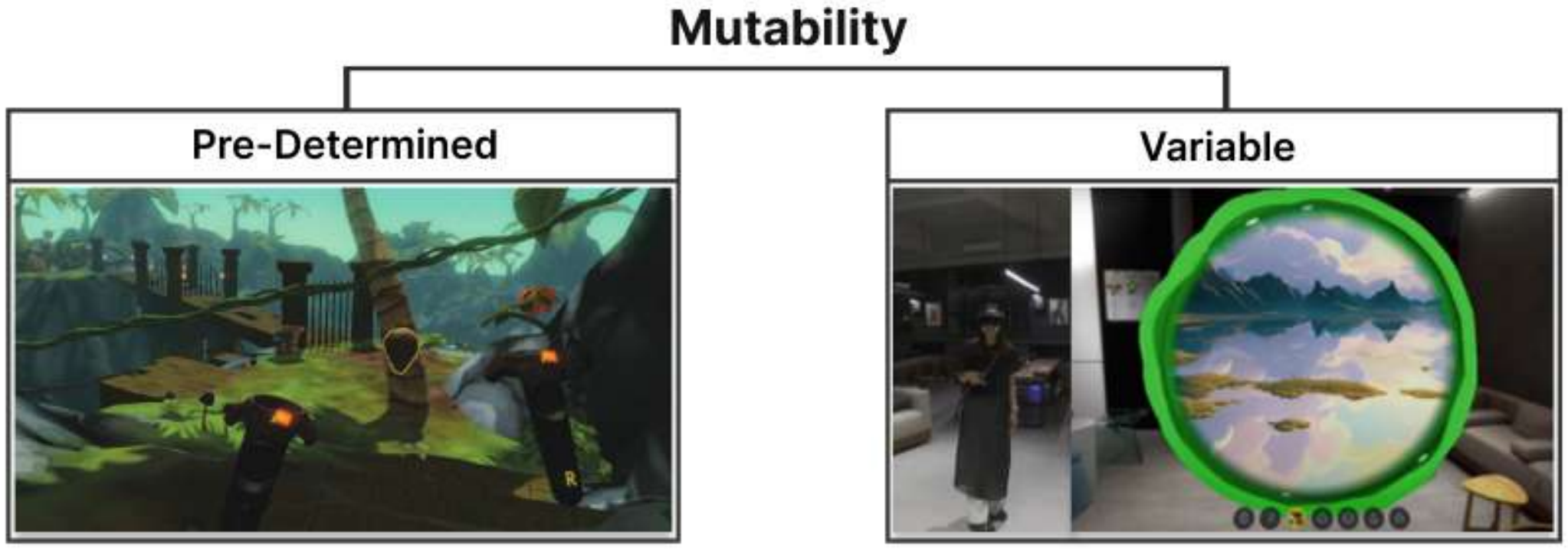

**Figure 13: Examples of affective XR systems illustrating the Mutability design dimension. Pre-determined: *Stressjam* [87] changes the colours of the virtual controllers that are pre-defined to indicate users' stress level from low to high (blue to orange), measured by HRV. Variable: Chang et al.'s systems [15] apply AI to generate virtual restorative environments for stress relief, measured by EEG. Each entry can generate a different virtual recovery environment.**

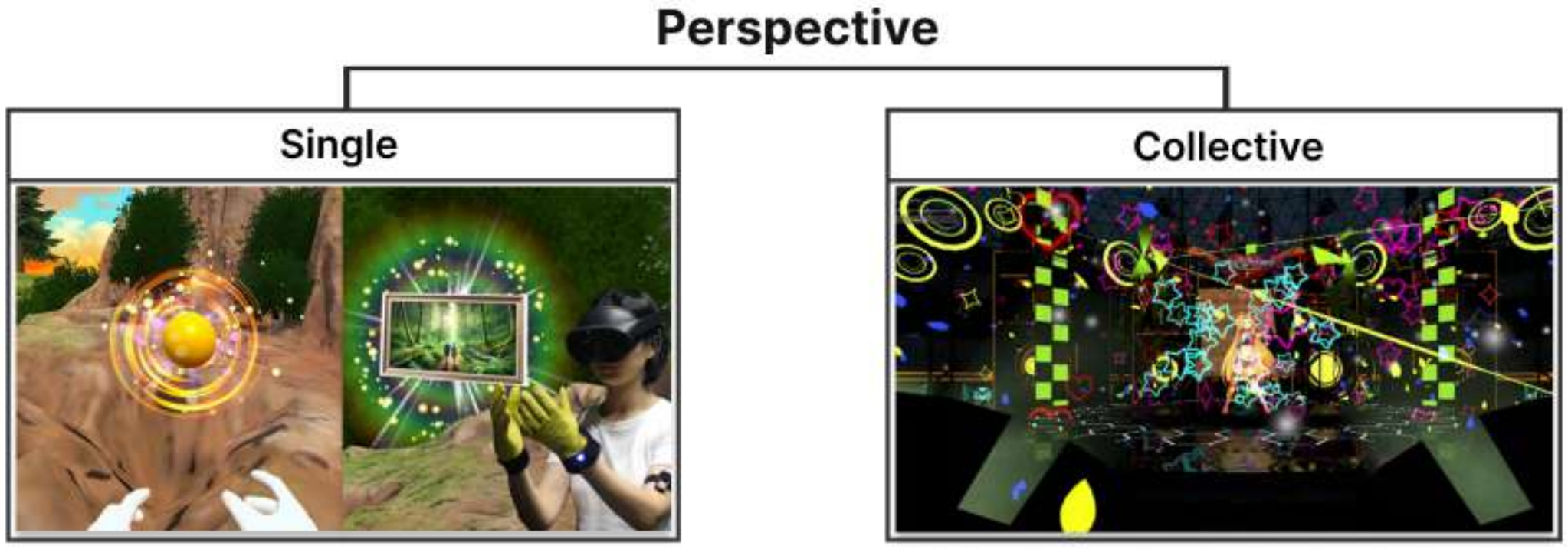

**Figure 14: Examples of affective XR systems illustrating the Perspective design dimension. Single: *Re-Touch* [47] shows autobiographical memory recall on affective state by monitoring PPG and EDA to adjust the haptic intensity. Collective: Muñoz-González et al.'s system [97] visualises the neurofeedback of multiple players in the music concert by different colours and forms of the visual effects (e.g. small stars, large stars, hearts, or cherry blossom petals).**

*Re-Touch*, which offers a VR experience for revisiting personal memories to enhance autobiographical recall through EDA by adjusting the haptic feedback intensity and the VR content [47]. Hybrid approaches incorporate both real-time and historical data; one paper visualised ECG data using a sequence of beating heart icons to simultaneously display both real-time and asynchronous physiological information in a virtual game, thereby supporting players in better understanding their in-game activities [55].

*5.2.9 Spatiality: Anchored vs. Virtual Co-moving vs. Real Co-moving.* (See examples in Figure 16). Our final dimension describes whether and how the emotion representation changes position within a virtual environment based on the user's perspective. Anchored representations ($N$=66) display biodata in a fixed spatial position, enabling consistent visibility within the user's view. Common placements include the left side of the visual field [60, 81], the right side [18, 73, 110, 152, 164], the center [24, 154], or the bottom-middle area [17, 28, 155].

Virtual co-moving representations ($N$=11) attach biodata visualisations to dynamic virtual references such as gaze points, or virtual elements. A number of systems implement virtual co-moving representations by linking biodata to dynamic virtual entities. These visualisations move in accordance with the user's gaze [65], virtual body [55], virtual hand [45, 47], virtual controllers [87], avatars [18, 51, 52, 125], or other objects in the virtual environment [107, 166]. Real co-moving representations ($N$=5) are similar but overlay biodata directly onto physical objects or human subjects, causing the emotional feedback to move along with real-world targets. Examples include augmentation onto a cube [37], a sandbox [120], a board [150], or around a person [130, 153] in the real world. In summary, a critical distinction between virtual and real co-moving representations is that the former anchors visualisations to virtual elements within a synthetic environment, while the latter overlays them onto physical objects in the real world.

## 5.3 Challenges for Emotion Understanding and Sharing

Our final results section presents four key challenges in designing representations of emotion, based on our analysis of the literature. We focus on interaction design challenges that impact the meaningful representation and interpretation of emotional states in affective XR, rather than exploring technical issues such as sensor interference [60, 80] or motion discomfort and fatigue [90, 133].

*5.3.1 Distrust and Inconsistent Feedback.* The reliability and accuracy of physiological feedback from sensing systems are crucial for enhancing user experience and understanding. For instance, users may have an inaccurate perception about their actual heart rate [25]. Some systems have faced challenges in maintaining consistency and fostering user trust. User perception itself introduces additional variability and is influenced by deceptive biofeedback. In one study that involved deceptive audio feedback about the user's heart rate, most participants did not notice the manipulation except in the most extreme cases [16]. One of their users expressed disappointment after encountering inaccurate feedback, stating, *"Once I found it was wrong, I didn't listen to him at all afterwards"*.

Content and stimulus design can also affect physiological responses. Seulki et al. suggest that the physiological effects of disturbing content in their system were not as strong as expected, leading to inconsistent biofeedback, possibly due to the physiological alteration being more resistant to the negative content (e.g. anxiety) than the positive one (e.g. calm) [93]. However, users virtually hold the panic-inducing stimulation of their own beating heart in their hand, supported by haptic and auditory feedback, which could be too intense for some patients [131]. Emotional exaggeration presents particular challenges for authenticity, one highlighted, *"Exaggerating my feelings in the exercise made me feel more understood by others"* when sharing partners' emotions by their own avatars in a cooperative bomb defusal game [51]. However, another noted, *"While it was interesting, it sometimes felt inauthentic to exaggerate emotions"* [51].

*5.3.2 Privacy Concerns and Discomfort.* The collection and visualisation of real-time physiological data can raise concerns about privacy, emotional vulnerability, and the potential for manipulation. This is because biodata is intimate and can reveal information about a person unintentionally. One study involving biofeedback in VR games found that privacy issues became a concern when heartbeat visualisations and movement data were shown to other players [55]. Privacy concerns were also highlighted in the study of *Empathic AuRea*, where participants paired with strangers reported higher levels of worry about disclosing physiological data compared to those paired with friends or acquaintances [153]. This transparency of emotional state, especially if perceived as intrusive or inaccurate, can lead to ethical concerns [30].

Additionally, there is a risk that using biodata can trigger undesirable affective experiences. One study found that visual changes in the colour of a virtual sky triggered 'anxious-like feelings' [113]. Another study found that showing biodata enhanced users' gaming experience but simultaneously made it 'incredibly stressful' [70]. Other studies involving AI revealed fears about emotional relationships with artificial entities. In a study by Chang et al. [16], participants noted that *"if the virtual human can understand and respond to my emotion like a real human, I am afraid I will emotionally bond to this virtual human"* [16]. Another added, *"... I would feel it's horrible if the agent is too accurate in detecting my emotional state. I have no privacy in front of such a virtual human ..."* [16].

*5.3.3 Distraction and Cognitive Overload.* Many systems involve biofeedback visualisations that are overlaid onto a virtual or physical environment in which the user is completing a task. While studies suggest these visualisations can be useful, their design must balance salience and subtlety, as they risk becoming distracting during periods of high cognitive load or intense interaction. For example, Dey et al. designed a system to show heart rate using simple heart icons in fast-paced gaming [24]. While some users expressed that the biofeedback enhanced their attention to a partner, others had the tendency to focus on the heart data over the gaming experience [24]. A similar effect was observed by Chen et al., who found that a heart rate visualisation attracted the attention of the user most of the time, even when it was not desired [17].

Another study suggested that processing too much information from biofeedback was challenging, and most exhibited minimal attention towards the physiological cues displayed by their partners [125]. Examples include *"the considerable use of colour was too distracting"* [153], *"too many visualisations were distracting and blocked the vision"* [55], and dynamic elements *"kind of distracted me because it was moving too fast"* [90]. In one study, participants mentioned distractions such as background noise or repetitive patterns that occasionally disrupted focus, while others reported occasional sensory overload where competing inputs disrupted focus [151].

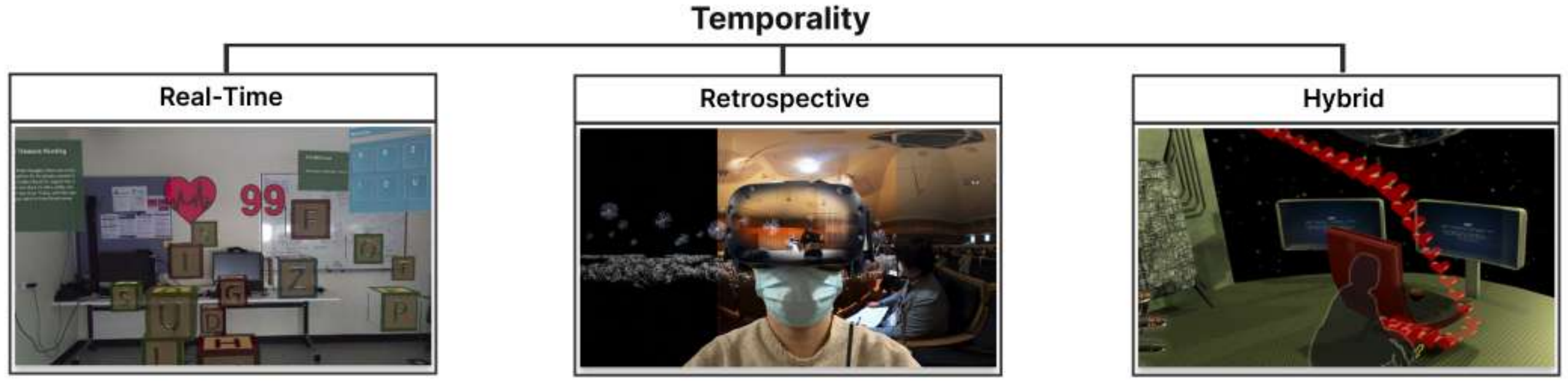

**Figure 15: Examples of affective XR systems illustrating the Temporality design dimension. Real-Time: Dey et al's system [24] displays real-time heart rate feedback with beats per minute (99) on an AR code decryption game. Retrospective: Meng et al.'s system [89] visualises BVP and EDA data of the audience from a recorded music concert to reproduce how a previous audience felt during a musical performance. Hybrid: *My Heart Will Go On* [55] visualises players' ECG data shown by both real-time and historical heart icons in a virtual escape game.**

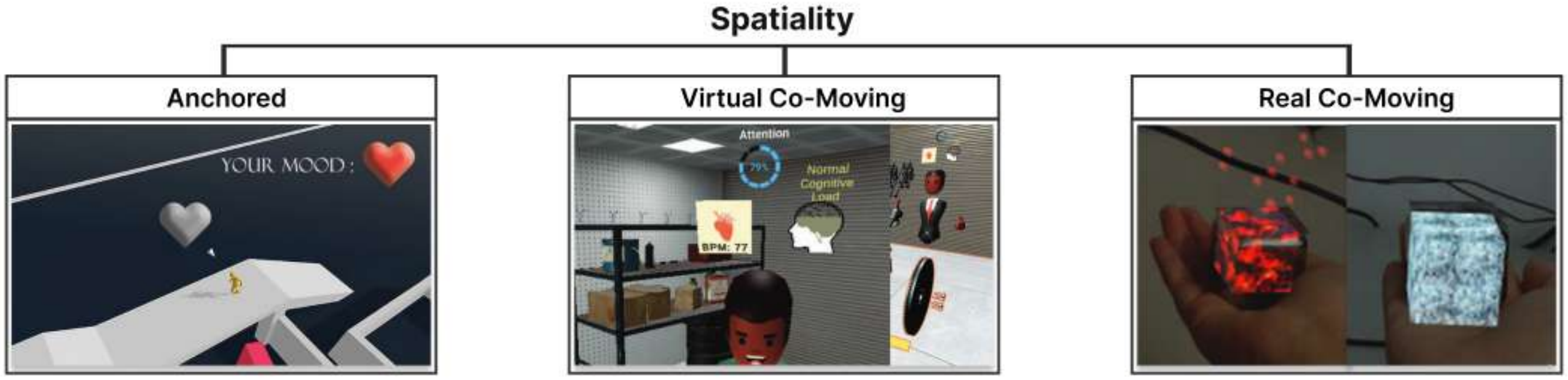

**Figure 16: Examples of affective XR systems illustrating the Spatiality design dimension. Anchored: *Tell Me How You Play* [18] indicates the player's mood by an anchored heart icon on the top right side in a VR Giants game, while a partner's mood is shown as a heart icon moving with the yellow avatar. Virtual Co-Moving: Sasikumar et al.'s system [125] visualises a collaborator's EEG (shown as 'Attention'), HR (shown as 'BPM'), and GSR (shown as 'Cognitive Load') on top of the avatar and moving along with the avatar when doing an engine assembly task. Real Co-Moving: Eom et al.'s system [37] augments the emotional response by overlaying different pictures on a real object based on arousal and valence, measured by EEG.**

Addressing the issue of distraction, some designs have successfully balanced engagement and subtlety, with one user from the *LifeTree* system noting that *"the blurry part of the game helped me understand if I am breathing rhythmically or not in a way that does not distract me from my breathing"* [104]. Nonetheless, approaches that aim to minimise distraction by making the feedback more subtle risk causing disengagement. A simple heart design created by Dey et al. led some participants to ignore or overlook the subtle visual HR cue [25]. Similarly, audio feedback proved difficult to perceive in stressful and action-packed environments, with one participant remarking, *"I am not sure whether I noticed the feedback as it was somehow hidden behind the sounds of gunshots"* in a VR shooting game [26].

#### 5.3.4 *Lack of Clarity and Interpretation Difficulty.*

The last challenge concerns the ability for users to understand and interpret the biosignals that they are being shown. Studies have highlighted several challenges related to interpretation. First, some systems involve multiple signals being shown at once, but combining visuals in this way can make for an overly complex experience. In one study, Sasikumar et al. showed participants visual representations of a work partner's heart rate, cognitive load, and attention [125]. Their participants emphasised the need for information to become overwhelming, with one stating, *"I would rather see everything first, then I can choose which is important for each task. Cognitive load tells me if my partner needs help... Heart rate tells me if my partner is excited... Attention tells me if he is focused".*

The choice between a literal versus a mapped representation of emotion can require users to decipher the meaning of the information, introducing analytical effort. One study involving literal representations found that participants were able to easily understand the meaning of a heartbeat visualisation, but a symbol representing ECG was 'too scientific', creating confusion about its meaning [55]. Conversely, systems that involve abstract representations sometimes create ambiguous meaning. This could lead to misalignments in understanding what a partner is feeling [130]. The lack of clear attentional guidance exacerbated these issues, with some users struggling to understand the meaning of those visualisations without explicit cues such as text saying 'look here' [55].

Lastly, metaphor selection and cultural specificity also emerged as influential in some studies. In one example, the visual metaphor used for anger (fire) was found to induce fear, suggesting the need for improved specificity [59]. Cultural factors further complicated interpretation, as differences in cultural norms for emotional expression can affect how users judge or perceive their biofeedback [85]. For example, the shape selection for visual effects was influenced by the Japanese concept of affective engineering, wherein most of them have a stronger reaction to specific shapes (e.g. hearts and cherry blossom petals) [97].

# 6 Discussion

Our scoping review provides an improved understanding of how XR systems have been designed to support emotion sharing based on biodata. In this section, we reflect on what our analysis reveals about this body of work and consider future directions for affective XR research.

## 6.1 The Current Landscape of Affective XR Systems

Our review reveals a diverse landscape of affective XR technologies, which aim to enhance affective experience through embodied bio-sensing and immersive feedback. Unlike traditional approaches that interpret physiological signals in screen-based or static contexts, XR enables users to inhabit and co-regulate with emotional representations, transforming emotions into multi-sensory feedback within simulated environments. Additionally, affective XR can create a dynamic loop where sensing, expressing and regulating co-evolve dynamically in a tailored virtual world. Our review illustrates how specific XR technologies have unique advantages for designing affective experiences. In the papers we analysed, VR systems were presented as having strong internal validity for emotion induction and regulation [81, 157, 168], exposure therapy [69, 131], or guided mindfulness [86, 93, 120], because of the fully controlled virtual environments where every cue can be designed to elicit or modulate affective states. In contrast, AR and MR systems can extend emotion into situated, real-world environments by overlaying affective cues onto physical surroundings [37, 130, 150, 153].

Our findings highlight a considerable and growing literature exploring the opportunities for XR systems to enable novel experiences centred around emotion. Focusing on the nexus of biodata, emotions, and XR, we have shown how the notion of affective XR can be a useful lens for analysing this literature as a coherent body of work. Interestingly, almost half of the papers were published in HCI venues such as CHI, IJHCS or DIS, showing our community's strong interest in the topic. Despite being scattered across multiple venues, we found that the literature is sufficiently coherent to reflect patterns of design intention, enabling us to outline a set of four key motivations and two taxonomies for reflecting on the design of these systems.

First, subsection 5.1 contributed an expanded taxonomy of how biodata sharing can be enabled in XR. Our taxonomy was based on an existing framework [95], elaborated into a set of ten approaches that show how biodata sharing can be enabled. This taxonomy reveals considerable diversity within the affective XR literature, but also highlights design opportunities. While many systems were intended for use by individuals, just under half of the technologies were designed to support emotion sharing between two or more users, and several biodata sharing setups had only been explored in a single paper (e.g. [28]). This reveals significant potential for applications involving interpersonal emotion sharing. We also highlighted how this sharing can be configured symmetrically or asymmetrically. Some interventions were designed to understand the effects of receiving information about a collaborative partner, whereas others aim to support mutual reflection around physiological data. These latter systems attempt to achieve emotional alignment between users through more complex designs that involve both individual and mutual modes of reflection [144].

Second, subsection 5.2 revealed nine design dimensions that are evinced by current systems in the literature. We emphasise that these dimensions are not mutually exclusive: rather, they are parameters along which a given design may vary. For example, a system might involve an *interactive*, *literal* biofeedback visualisation of *real-time* data, whereas another might also be *interactive* but show *mapped* and *retrospective* biofeedback. Our analysis highlights that some opportunities within these dimensions have been well-explored. For example, the majority of systems include *non-interactive* visualisations (81/82), using *passive* representations of emotion (80/82) that show *real-time* data (75/82). Conversely, several opportunities remain largely unexplored. These include *deceptive* designs (5/82), those which employ *variable* data displays (5/82), *collective* representations of emotion (3/82), and *interactive* data displays (1/82). We anticipate that these dimensions will be useful for researchers when considering opportunities to expand research on affective XR, especially given the somewhat lopsided exploration of several dimensions within the literature.

In addition, there are opportunities to better understand overlaps between these dimensions and explore underused areas of the design space. For example, one frequent overlap is between *passive*, *real-time*, *truthful*, and *non-interactive* data representations, likely because this combination of features is a straightforward way of designing a biodata visualisation. Conversely, there are areas in which design parameters overlap less frequently, and thus we know little about the benefits of combining them. Examples of rare overlaps in the corpus include the use *variable* represents with *deceptive* biofeedback [16] and *proactive* guidance from an agent for perceiving empathy [48]. Another rare overlap is between *variable* systems that use AI to generate virtual environments on *retrospective* data [47] and *collective* representations to raise shared emotional awareness among users [30].

Taking our taxonomies together, we anticipate that they will be useful for guiding future design work towards areas of opportunity that remain underexplored. They may also be useful for thinking through design choices that arise as tensions between technical capabilities and user needs. These may come into play when addressing the challenges identified in our review (see subsection 5.3). For example, biofeedback visualisations must be salient enough to inform, yet subtle enough to avoid disrupting primary tasks or inducing cognitive overload, particularly in fast-paced or stressful scenarios [24, 28, 100]. Our taxonomies can support researchers in thinking through issues that may arise along particular design dimensions. As just one example, the trade-off between abstraction and literalness affects interpretability: overly abstract visualisations risk misinterpretation and cognitive overload [22, 35, 113, 130] while excessively literal ones may reduce engagement [80, 125].

These considerations suggest that the efficacy of affective XR systems may be contingent on their ability to navigate the fundamental tension between providing meaningful, actionable biofeedback and maintaining immersion by avoiding disruptive or cognitively taxing interactions. Navigating these tensions is essential for leveraging the benefits of biodata-driven interaction while mitigating key experiential challenges.

## 6.2 Future Research Avenues

*6.2.1 Expanding the Design Space of Affective XR.* Our review highlights that the majority of existing affective XR systems focus on VR, while AR and MR were less explored. In addition, passthrough AR and augmented virtuality (AV) were completely absent from the corpus, representing opportunities for future design work. Passthrough AR in particular will likely make it easier to create affective XR experiences by providing a wider field of view than some MR headsets, and by allowing users to transition between AR and VR environments.

As highlighted above, our analysis identified several dimensions that are underused in affective XR design. Figure 17 positions the 6 most-underused dimensions against the goals that systems using them have aimed to achieve, and the challenges that arose when these design parameters were employed. In terms of goals, systems for emotional awareness and understanding have employed *variable* and *interactive* designs [47, 151] to make emotions visible. The remaining four dimensions (Hybrid, Deceptive, Proactive and Collective) have been unexplored. The second goal, emotion regulation and induction, has been pursued across these underused dimensions more frequently [15, 26, 111, 150]. The remaining two goals have only been explored through hybrid designs, collective representations, and variable visualisations. Future research could extend our framework by systematically analysing the interdependencies and interrelationships between the dimensions, which could further investigate their relative importance and combined effects on affective XR.

Regarding challenges, privacy concerns are prominent, as these systems collect the users' biodata and often explicitly display internal states (e.g. arrows for stress levels [16], voice reporting physiological changes [48]), which makes users feel exposed. A *deceptive* and *variable* system may reduce perceived reliability [16]. Furthermore, multiple concurrent stimuli, such as *hybrid* biodata visualisations, *collective* biodata representations, and task demands, could significantly increase cognitive load [48]. While *interactive* and *hybrid* designs can enrich user experience, they may exacerbate interpretation difficulty if not intuitively designed to combine multiple cues [55, 151]. Given that fewer than twenty such systems were identified as using underused dimensions in total, we suggest that future research should not only explore these underused dimensions but also anticipate and mitigate associated challenges through clearer feedback mechanisms and deliberate attention to cognitive and ethical boundaries. By consciously aligning dimensional exploration with both goals and constraints, the next generation of affective XR systems can achieve their objectives while addressing key concerns.

*6.2.2 Exploring Multimodal Feedback and Personalisation.* Our findings highlight that current affective XR systems predominantly rely on visual feedback, with auditory and haptic modalities used less often as the primary means of biofeedback. Multimodal feedback, such as the integration of visual and auditory information, could help to mitigate the challenge of distraction and cognitive overload [97]. This is because using non-visual cues such as audio or haptics for emotion communication can reduce visual distraction while preserving immersion and empathy [35]. Dynamically adapting challenge levels or environmental elements based on real-time biodata may also help to sustain engagement without over-distraction. Here, such adaptation itself could become an implicit feedback mechanism, reducing reliance on only explicit visualisations of emotional status.

Future studies might investigate how multimodal feedback designs might usefully contribute to emotion conveyance, particularly personalised haptic interfaces that can communicate nuanced emotional states through tactile cues [e.g. 128]. Additionally, although olfactory and gustatory modalities remain beyond the scope of current affective XR systems, there is potential for future research to explore their integration for richer, more immersive emotional experiences [e.g. 39]. Emotion is often a complex experience, involving not just changes in internal state but also facial expressions, physiology, and movement, e.g. shaking in response to fear. Creative design work might attempt to explore open questions that fall at the intersection of emotion and different sensory experiences. For example, what does happiness 'feel' like? And how might we 'see' or 'touch' fear from a partner via XR technology?

Personalisation may also help to address challenges that arise in affective XR. Users vary in their perceptual preferences [128], gender [7], and cultural backgrounds [66, 97, 104, 123], which could influence how they interpret literal versus mapped emotion representations. Furthermore, overly literal representations could feel clinical and increase cognitive load [80, 125], while overly mapped representations risk being misunderstood [22, 35, 113, 130]. Some prefer seeing direct raw data for self-reflection on their emotions [125], while others desire interpreted emotional insights by the system [16]. We therefore suggest exploring personalised multimodal feedback, such as metaphoric haptic patterns or culturally resonant audio cues, which can convey nuanced emotional states without intruding on the user's primary task.

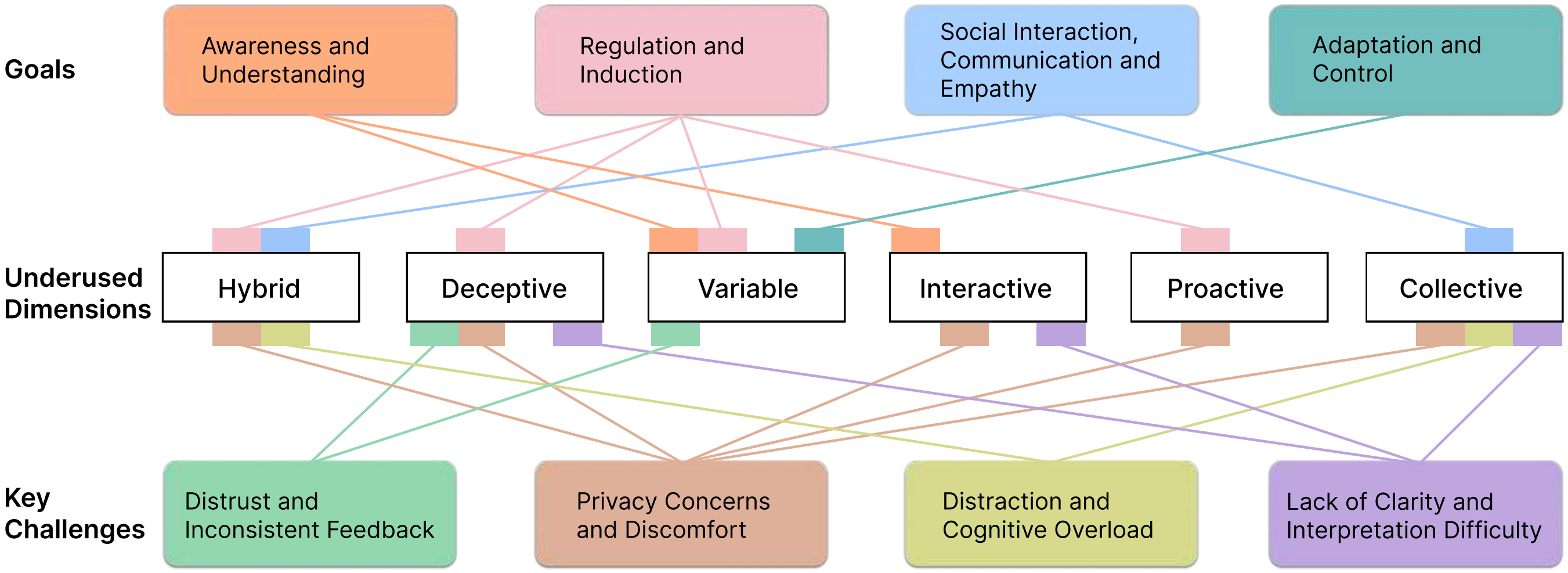

**Figure 17: The six most underused dimensions against the goals that systems have aimed to achieve, and the challenges that arose when these design parameters were employed, based on our interpretation and analysis from the final corpus.**

*6.2.3 Investigating AI-supported Affective Experiences.* Aligning with the rapid advancement of AI technologies, the integration of AI presents a significant opportunity for future work on affective XR. However, current research is still in its infancy; only 4 papers in our corpus used AI, and they were all published in the past two years [15, 16, 48, 99]. Two key opportunities are the nascent exploration of (1) using emotion-aware agents to support interaction, and (2) using AI-Generated Content (AIGC) to build an affective XR world. Empathic agents which recognise users' physiological state can support empathy and social presence in virtual environments by acting as a 'social companion' [16, 48]. Such agents could also adapt their behaviour by providing encouragement or speaking more slowly when the user is stressed [99]. AIGC can facilitate the adaptation of virtual elements or environments, compared to pre-determined assets [15]. Here, we call for more research on affective XR systems that are not static environments that simply monitor users, but instead use proactive, collaborative AI partners that intelligently reason about, respond to, and shape the user's emotional journey.

Incorporating AI, however, could introduce critical challenges related to ethics, privacy, safety, autonomy, and trustworthiness. Future studies must therefore embed ethical considerations into the design of AI-driven affective XR, ensuring transparency in biodata usage, securing informed consent for sensitive biodata processing, and preserving user control over automated interventions to avoid manipulative or overly intrusive experiences. This also led us to not only try to clarify and interpret emotional states to the researchers, but also make it transparent to all the users in their user studies, including authentic and deceptive biodata.

## 7 Conclusion

In this paper, we have contributed a scoping review of 82 publications that enable us to map the current landscape of the literature on affective XR technologies. Our analysis reveals growing interest in affective XR within the HCI community. We identified commonly used types of biodata, along with specific affective states that systems have been designed to convey, alongside their applications across various XR technologies. Our review also reveals a set of common goals for this work, including emotion awareness and understanding, emotion regulation and induction, emotions for social interaction, communication and empathy, and emotions for adaptation and control. By analysing the existing landscape of system designs, we provide an improved understanding of biodata flows within affective XR, and design dimensions that characterise different ways of representing emotions. We finally identified four key challenges for emotion sharing, based on the interpretation of findings from our corpus. By mapping these, our review reveals a heterogeneous body of work with significant potential for further research, including expanding the scope of technologies used, the interaction techniques and modalities employed, and the potential for AI integration in affective XR. Collectively, we contribute to the growing interest on emotion within the HCI community [156] and to recent conversations on the use of biodata as input for interactive systems that enable understanding of ourselves and others [20, 94].

## Acknowledgments

This work is funded by an RMIT Vice-Chancellor's PhD Scholarship (VCPS).

# A Search String

**Table 3: Search String. The "AND" operator was used between different keyword sets, whereas the "OR" operator was used within the same keyword set.**

`("biometric*" OR "physiologic*" OR "biosignal*" OR "biodata" OR "biosens*" OR "bio-responsive" OR "biofeedback" OR "neurofeedback" OR "bio-signal*" OR "bio-sens*" OR "brain-computer" OR "BCI" OR "EEG" OR "electroencephalogra*" OR "brain activity" OR "heart rate" OR "ECG" OR "HRV" OR "electrocardiogra*" OR "cardiac activity" OR "PPG" OR "breath*" OR "respiration" OR "electrodermal" OR "EDA" OR "GSR" OR "galvanic skin response" OR "skin conductance response" OR "sweat" OR "skin temperature" OR "electromyogra*" OR "EMG" OR "eye tracking" OR "EOG" OR "electrooculogra*")` **AND** `("emotion*" OR "affect*" OR "empathic" OR "feeling*" OR "mood")` **AND** `("extended reality" OR "XR" OR "mixed reality" OR "MR" OR "augmented reality" OR "AR" OR "virtual reality" OR "VR" OR "head-mounted display" OR "head-up display" OR "head-worn display" OR "headset*" OR "HMD" OR "immersive environment*" OR "virtual environment*" OR "virtual space*")`

# B Exclusion Criteria

**Table 4: Exclusion Criteria (EC) to exclude irrelevant articles during the PRISMA process. Each EC is defined as described.**

| # | Criteria | Description | Count |
|---|---|---|---|
| **EC1** | Does not Focus on Emotion | The article uses "affect", but as a verb to mean "influence"; Or, uses physiological signals in XR but not for emotion-related reasons | 259 |
| **EC2** | XR Technologies | The article mentions immersive environments but does not use XR technology | 53 |
| **EC3** | Biodata Integration | Did not use at least one kind of biodata in their system or methodology | 21 |
| **EC4** | Not Target Content Type | Article was not a peer-reviewed conference paper, short paper (e.g. late-breaking work), journal article, or book chapter | 16 |
| **EC5** | Overlapping Publication | The article has overlapping results with another paper about the same project or system | 85 |
| **EC6** | Dataset | The article only provides a dataset, without an accompanying system or study | 24 |
| **EC7** | Models | The article focuses only on improving machine learning models to recognise emotion | 181 |
| **EC8** | Review | The article is a literature review or survey publication | 42 |
| **EC9** | No Biofeedback | The article reports a study using biodata, but does not involve biofeedback or only uses biodata as a study measurement, e.g. dependent variable | 428 |

# C Data Extraction Template

**Table 5: Data Extraction Template. Multiple sub-questions could be asked for subsequent data analysis within a single Coding.**

| Source Characteristics | Participant Information | Immersive Settings | User Study |
|---|---|---|---|
| C1 General information | *C5* Total number | *C8* Immersive type | *C14* Study design (e.g. task) |
| C2 Venue and year | *C6* Age (e.g. M, SD) | *C9* Hardware and software | *C15* User (e.g. number, roles) |
| C3 Research method | *C7* Gender | *C10* Biodata collected | *C16* Study measurement |
| C4 Objectives and findings | | *C11* Specific emotion captured | *C17* Study dataset used |
| | | *C12* Emotion model followed | *C18* Participant's comments |
| | | *C13* Emotion representation (e.g. visual, audio, haptic) | *C19* Interaction design (e.g. symmetry, synchronous) |

## D Open Datasets Used in Affective XR Studies

**Table 6: Open dataset collected from EC6 and final corpus. *Name of dataset from the publication with over fifty citations; **Name of dataset from the publication with over a hundred citations. Citations last accessed on 22nd Aug, 2025.**

| Name | Description | Link | Ref |
|---|---|---|---|
| *CEAP-360VR | A multimodal dataset from 32 participants watching eight 360° VR videos, collecting continuous valence-arousal ratings, physiological signals, and head/eye movements | https://github.com/cwi-dis/CEAP-360VR-Dataset | [163] |
| Data Descriptor | A dataset of physiological signals from 24 participants in traditional and partially immersive learning scenarios on a Humanities topic, including EEG, EDA, BVP, and psychometric data | https://doi.org/10.6084/m9.figshre.24777084 | [119] |
| **DEAP | A public database for emotion analysis using physiological Signals (EEG, etc.) from 32 participants watching 40 one-minute-long music videos | https://doi.org/10.1109/T-AFFC.2011.15 | [75] |
| **DREAMER | A public EEG and ECG dataset from 23 participants watching music videos to elicit different emotions | https://doi.org/10.1109/JBHI.2017.2688239 | [72] |
| EEVR | A dataset from 37 participants featuring physiological signals (EDA, PPG) and paired raw textual descriptions of emotions elicited by 360-degree VR videos | https://openreview.net/forum?id=qgzdGyQcDt | [136] |
| Emo-DB | A German emotional speech dataset of 535 utterances from actors, used to classify seven emotions including anger, joy, sadness, and fear | https://www.kaggle.com/datasets/piyushagni5/berlin-database-of-emotional-speech-emodb | [77] |
| EmoNeuroDB | An EEG dataset from 40 individuals mimicking an avatar's facial expressions in VR to elicit six emotions: fear, joy, anger, sadness, disgust, and surprise | https://voxellab.pl/EmoNeuroDB/ | [34] |
| FilmStim | A database of emotion-eliciting film clips that represent emotions such as sadness or anger | https://doi.org/10.1080/02699930903274322 | [153] |
| GAMEEMO | A dataset of EEG signals from 28 participants playing four different computer games (boring, calm, horror, funny) to recognise emotions based on aural and visual stimuli | https://doi.org/10.17632/b3pn4kwpmn.3 | [92] |
| IADS | A database includes validated, labelled, and emotionally-evocative international affective digitised sound stimuli related to anger, fear, joy, and sadness | https://psycnet.apa.org/record/2020-13958-004 | [59] |
| IAPS | A international affective picture system contains images rated for valence and arousal, evaluated to elicit anger, fear, joy, and sadness | https://doi.org/10.1037/t66667-000 | [59] |
| MUMTG | A dataset from 39 participants performing guided assembly tasks with AR or monitor instructions, including head/eye tracking, HR, GSR, VR images | https://doi.org/10.17632/b7c2h6cbc6.1 | [53] |
| OASIS | A open affective standardised image set of pictures that provides normative arousal and valence ratings | https://doi.org/10.3758/s13428-016-0715-3 | [37] |
| PEM360 | A dataset of head movements and gaze recordings in 360° videos, with self-reported emotional ratings and physiological measurements (EDA, heart rate) | https://gitlab.com/PEM360/PEM360 | [116] |
| *RAGA | A physiological dataset from 33 participants playing racing games in VR and on a monitor, tracking ECG, EMG, and EDA to predict arousal and valence | https://github.com/grano00/GameVRRacingPhysioDB | [43] |
| SEED | An EEG dataset containing brainwave data from participants watching emotive film clips, used for emotion recognition | http://bcmi.sjtu.edu.cn/~seed/seed.html | [62] |
| VRAT | A multimodal dataset from 25 participants using two VR games to induce calm and stress, tracking ECG and RSP signals to classify affective states | https://gitlab.com/hilabmsu/database-vrat | [68] |
| **VREED | A dataset from 34 participants to elicit emotions using 360° VR videos, containing self-reported assessments, ECG, and GSR physiological data | https://www.kaggle.com/dataset/e9e93acd547401db81cd9988c96760d823142fa894a4a79c6af971949d4bdf85 | [145] |
| VREEG | An EEG emotions dataset from 70 participants in a VR environment, categorised into a 3-class (negative, neutral, positive) and a 4-class (happy, sad, neutral, fear) subset | https://github.com/vreegemotions/VREEG_Datasets | [3] |
| VRMN-bD | A multi-modal dataset collected from 23 players in VR horror games to analyse fear responses, including posture, audio, and physiological signals | https://github.com/KindOPSTAR/VRMN-bD | [167] |

# E Measures Used in Affective XR Studies

**Table 7: Measurement of emotion and other metrics used in the final corpus. Some studies could involve multiple measurements.**

| Measurement | Abbr. | N | Papers |
|---|---|---|---|
| **Emotion** | | **66** | |
| *Self-Assessment Manikin* | *SAM* | *15* | [17, 18, 25–27, 29, 37, 48, 59, 70, 82, 89, 90, 153, 164] |
| *Positive and Negative Affect Schedule* | *PANAS* | *13* | [17, 24–28, 48, 86, 126, 137, 148, 154, 168] |
| *Customised Questionnaires* | - | *12* | [7, 9, 16, 21, 23, 85, 97, 111, 123, 133, 144, 166] |
| *State-Trait Anxiety Inventory* | *STAI* | *8* | [15, 73, 93, 113, 120, 131, 154, 157] |
| *Perceived Stress Scale* | *PSS* | *2* | [9, 135] |
| *Visual Analog Scale* | *VAS* | *1* | [157, 161] |
| *Beck Anxiety Inventory* | *BAI* | *1* | [131] |
| *Brunel Mood Scale* | *BRUMS* | *1* | [155] |
| *Center for Epidemiologic Studies-Depression* | *CES-D* | *1* | [151] |
| *Facial Affective Scale* | *FAS* | *1* | [73] |
| *Four Fs of Trauma (Fight, Flight, Freeze, Fawn)* | *Fs* | *1* | [73] |
| *Geriatric Depression Scale* | *GDS* | *1* | [118] |
| *Montgomery-Åsberg Depression Rating Scale* | *MADRS* | *1* | [131] |
| *Multidimensional Mood Questionnaire* | *MDBF* | *1* | [118] |
| *PAD emotional state (Pleasure, Arousal, and Dominance)* | *PAD* | *1* | [110] |
| *Physical Activity Affect Scale* | *PAAS* | *1* | [73] |
| *Profile of Emotional Competence* | *PEC* | *1* | [130] |
| *Short Stress State Questionnaire* | *SSSQ* | *1* | [153] |
| *State-Trait Cheerfulness Inventory* | *STCI* | *1* | [148] |
| *Stress Mindset Measure* | *SMM* | *1* | [87] |
| **Experience** | | **9** | |
| *Game Experience Questionnaire* | *GEQ* | *5* | [48, 61, 104, 149, 153] |
| *Experience Sampling Method* | *ESM* | *1* | [110] |
| *Flow Short Scale* | *FSS* | *1* | [168] |
| *Player Experience Inventory* | *PXI* | *1* | [153] |
| *System Usability Scale* | *SUS* | *1* | [35] |
| **Cognitive or Task Load** | | **6** | |
| *Task Load Index* | *NASA-TLX* | *5* | [9, 15, 48, 125, 153] |
| *Mind Indexes* | *MIs* | *1* | [7] |
| **Inclusion** | | **6** | |
| *Inclusion of Other in the Self* | *IOS* | *3* | [24, 27, 28] |
| *Social Connectedness Scale* | *SCS* | *2* | [151, 153] |
| *General Belongingness Scale* | *GBS* | *1* | [151] |
| **Mindfulness** | | **3** | |
| *Five Facet Mindfulness Questionnaire* | *FFMQ* | *2* | [120, 151] |
| *Toronto Mindfulness Scale* | *TMS* | *1* | [120] |
| **Presence** | | **2** | |
| *Networked Minds Social Presence Measure* | *NMSPM* | *1* | [124] |
| *Social Presence Questionnaire* | *SPQ* | *1* | [27] |
| **Mental Health** | | **2** | |
| *General Health Questionnaire* | *GHQ* | *1* | [73] |
| *Warwick Edinburgh Mental Well-being Scale* | *WEMWBS* | *1* | [157] |
| **Intention** | | **1** | |
| *Future Usage Intention Scale* | *FUSI* | *1* | [168] |
| **Perception** | | **1** | |
| *Godspeed Questionnaire Series* | *GQS* | *1* | [16] |

# F Screenshots of All Affective XR Systems

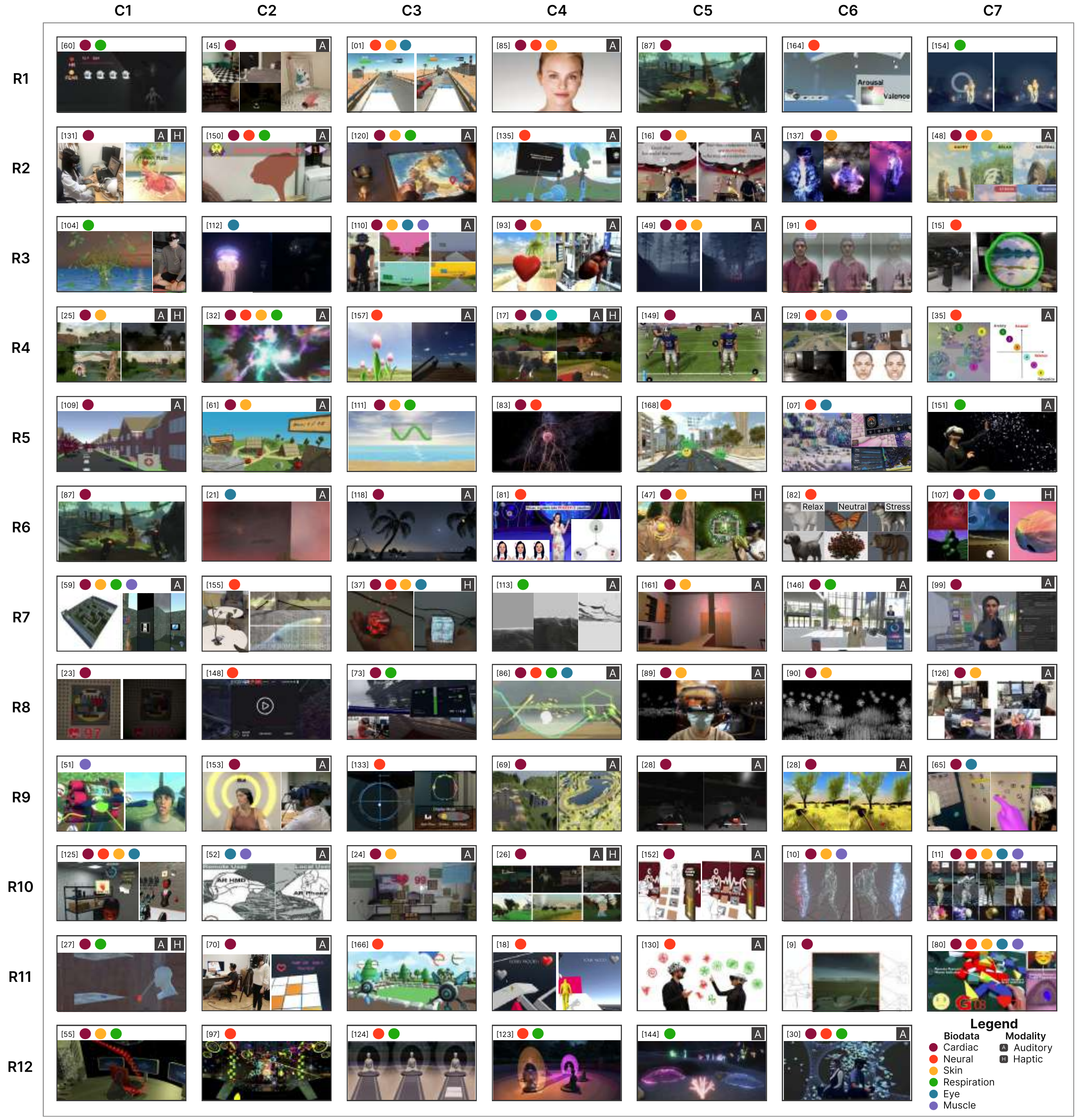

**Figure 18: Schematic representation of all systems sorted by subsection 5.1. Each category starts from left to right, top to bottom [$R_{start}$, $C_{start}$ - $R_{end}$, $C_{end}$]. Single-user: ① I feel mine [1,1-8,4]. Multi-user Asymmetry: ② I feel yours [8,5-9,1]; ③ You feel mine [9,2-9,3]; ④ We feel yours [9,4]; ⑤ We feel mine [9,5]. Multi-user Symmetry: ⑥ I feel yours, you feel mine [9,6-11,2]; ⑦ We feel each other [11,3-12,1]; ⑧ We feel ours aggregated [12,2]; ⑨ I feel yours, you feel mine, and we feel ours aggregated [12,3-12,4]; ⑩ We feel each other, and ours aggregated [12,5-12,6] (Nb. The total exceeds 82 as ④ and ⑤ are the same system from different users).**

## G Emotion Representations in Affective XR Studies

**Table 8: Summary of Emotion Representations by Modalities. The sum of visual representations is 82, while the sum of auditory and haptic representations is 42 and 8, respectively, since some systems did not involve these modalities.**

| Representation Type | N | Papers |
|---|---|---|
| **VISUAL** | | |
| **Environments** | **27** | |
| *Spatiotemporal objects (e.g. geometric primitives)* | *7* | [15, 35, 47, 89, 90, 130, 151] |
| *Whole environment changing* | *6* | [7, 30, 59, 97, 118, 135] |
| *Preternatural transmission (e.g. universe)* | *5* | [22, 32, 79, 91, 113] |
| *Lighting effects (e.g. darkening vignette)* | *4* | [21, 23, 49, 161] |
| *Virtual elements on top of real objects* | *3* | [37, 120, 150] |
| *Pre-set environments* | *2* | [25, 26] |
| **Symbols** | **15** | |
| *Icons* | *10* | [17, 24, 27, 28, 55, 60, 80, 93, 125, 154] |
| *Progress bar* | *5* | [65, 73, 110, 152, 166] |
| **Avatar** | **12** | |
| *Changing velocity, shape, and color* | *9* | [10, 11, 16, 18, 83, 123, 124, 126, 137, 159] |
| *Changing facial expression* | *3* | [51, 52, 85] |
| **Adaptive Settings** | **12** | |
| *Task or game difficulties* | *10* | [1, 9, 29, 45, 61, 69, 70, 109, 149, 168] |
| *Changing agent behavior* | *2* | [99, 146] |
| **Dynamic Motion** | **11** | |
| *Moving dot in quadrant* | *4* | [81, 107, 133, 164] |
| *Line going up and down* | *4* | [86, 111, 148, 155] |
| *Color changing* | *3* | [48, 87, 153] |
| **Creature** | **5** | |
| *Growing up* | *3* | [104, 144, 157] |
| *Moving position* | *1* | [112] |
| *Replacing with other creature* | *1* | [82] |
| **AUDITORY** | | |
| **Anthropomorphic** | **19** | |
| *Heartbeat* | *12* | [17, 21, 24–28, 70, 89, 93, 131, 161] |
| *Voice from speech* | *4* | [16, 48, 52, 85] |
| *Breathing simulation* | *3* | [104, 109, 113] |
| **Atmospheric** | **14** | |
| *Nature sounds (e.g. birds, insects, water, fauna)* | *8* | [30, 69, 118, 120, 144, 151, 152, 157] |
| *Background music atmosphere* | *4* | [49, 59, 61, 135] |
| *Instruments sounds* | *2* | [126, 150] |
| **Functional** | **9** | |
| *Tension control* | *5* | [35, 45, 86, 146, 149] |
| *Vibrating sound* | *4* | [32, 110, 130, 153] |
| **HAPTIC** | | |
| *Heartbeat Simulation* | *5* | [17, 25–27, 131] |
| *Others* | *3* | [37, 47, 107] |